# Annealing-Induced Magnetic Modulation in Co- and Y-doped $CeO_2$: Insights from Experiments and DFT

Hemant Arora[1], Atul Bandyopadhyay[2,*], and Arup Samanta[1,3,*]

[1]Quantum/Nano Science and Technology Laboratory, Department of Physics, Indian Institute of Technology Roorkee, Roorkee-247667, Uttarakhand, India

[2]Department of Physics, University of Gour Banga, Mokdumpur, Malda, West Bengal 732103, India

[3]Centre for Nanotechnology, Indian Institute of Technology Roorkee, Roorkee-247667, Uttarakhand, India

*Corresponding author. Email address: atulbandyopadhyay@yahoo.com; arup.samanta@ph.iitr.ac.in

**Abstract:** The potential applications of dilute magnetic oxides (DMOs) in magneto-optic and spintronic devices have attracted significant attention, although understanding their magnetic behavior is complex due to intricate interactions of intrinsic defects. The present study aims to investigate the effect of different annealing environments on the magnetic properties of polycrystalline transition metal cation (Co and Y) doped $CeO_2$ DMO with a 5% doping concentration of transition metal (TM). The objective is to investigate the defect interactions within the lattice through a comprehensive investigation involving structural characterizations, magnetic measurements, and first principle calculations. The results show that the $Ar/H_2$ annealing environment induced more oxygen vacancies than air-annealed samples. Consequently, field-dependent magnetization measurements revealed above-room-temperature ferromagnetism (RTFM) in both un-doped and TM-doped $CeO_2$. The ferromagnetic (FM) properties of $CeO_2$ resulted from carrier-trapped vacancy centers facilitating exchange interactions between the spins of magnetic ions. The Langevin field profile indicated that TM-doped $CeO_2$ formed more bound magnetic polarons (BMPs) during annealing in an $Ar/H_2$ environment, which contributed to the enhanced ferromagnetism. Similarly, enhancement in the magnetic properties with increasing oxygen vacancies is observed through first principle calculations. This suggests the potential for optimizing the magnetic properties of DMOs through controlled annealing processes.

## Introduction

The origin of spintronics, motivated by the growing demand for high-speed processing and extensive data storage capabilities, has guided a transformative era in current industries. Spintronics possesses the capability to integrate the unique characteristics of information carriers, such as charge carriers used in data processing and storage, and photons used in data transmission.[1,2] This combination offers the fascinating prospect of manipulating electrons not only through their charges but also by influencing their spins. Essential to the advancement of any technology is the proper selection of suitable materials. In the domain of spintronics, dilute magnetic semiconductors (DMSs) and dilute magnetic oxides (DMOs) have gained substantial attention due to their remarkable ability to combine the distinctive properties of both ferromagnets and semiconductors. The origin of magnetism in semiconductor families (II−VI, III−V, IV-IV, and IV−VI)[3–6] and oxides ($ZrO_2$, $CeO_2$, $SnO_2$, $Gd_2O_3$, etc.)[7–11] involves the strategic substitution of various impurities. This substitution involves TMs, rare-earth elements, and non-magnetic dopants like C, N, etc., collectively referred to as the ferromagnetic phantom '$d^0$'.[12,13] This substitution induces changes in the density of electrons and hole carriers density, leading to the rise of localized magnetic moments. These moments introduce magnetism to the system, paving the way for innovative applications in spintronics and magnetic-related technologies. DMOs possess several advantages over DMSs, including chemical stability, a wide bandgap, control over oxygen vacancies, catalytic properties, intrinsic stability, versatility in doping, and notably above-room-temperature ferromagnetism (RTFM). New approaches involving the engineering of DMOs structure, composition, and grain size offer avenues to enhance their functionalities, making them promising materials in various applications.

Cerium dioxide ($CeO_2$) among all the DMOs, has attracted considerable attention in the field of microelectronics, surpassing materials like $HfO_2$ and $Ta_2O_5$. This is primarily due to its excellent lattice match with silicon, exceptional insulating properties, and impressive chemical stability. $CeO_2$ also possesses remarkable dielectric properties ($\varepsilon \approx 26$), which allow for higher charge storage capacity, an essential feature for developing spintronic devices that require both charge and spin manipulation. Thin films of $CeO_2$ have also shown excellent performance in metal-insulator-semiconductor structures, making them a promising candidate for next-generation electronic applications.[14] Beyond its dielectric properties, $CeO_2$ is increasingly recognized for its potential magnetic properties, both in un-doped and doped systems. These magnetic properties can be harnessed to synthesize materials capable of generating spin-polarized currents, which is crucial for the manipulation of spin currents in spintronic devices. As a result, $CeO_2$ has become a focal point in the domain of silicon electronics, with ongoing research aiming to bring this material system for commercial use.

A key area of interest has been the observation of RTFM in un-doped $CeO_2$, which is often attributed to the $d^0$ effect. The $d^0$ effect occurs when metal oxides, especially rare-earth or transition metal oxides, exhibit magnetism due to the presence of oxygen vacancies.[15–19] These vacancies lead to the formation of localized electronic states, frequently involving transition metal ions, which can contribute to the observed magnetic behavior. The doping of TM ions in $CeO_2$ has become a widely studied approach for tailoring its magnetic properties. TM-doped $CeO_2$ has found applications in spintronics, catalysis, and other technological fields. While many studies have explored the origin of RTFM in doped $CeO_2$, the underlying mechanism remains highly debated. Some researchers argue that the source of RTFM in doped $CeO_2$ is the formation of dopant clusters,[20,21] while others propose that oxygen vacancies ($V_O$) are the key contributors to the observed magnetic properties.[22,23]

The studies that indicate the presence of TM clusters cite the strong FM behavior observed in these materials. Additionally, the field-cooled and zero-field cooled curves of these materials do not superimpose even at temperatures as high as 300 K, suggesting that the material is inhomogeneous and that the magnetic properties are largely due to the TM magnetic clusters within a nonmagnetic or paramagnetic matrix. On the other hand, studies attributing the origin of magnetism to oxygen vacancy formation report relatively small magnetic moments, even with high doping concentrations.[24–31] However, some studies report high magnetic moments without offering clear evidence of impurity effects, adding further complexity to the discussion.[32]

Given these ongoing debates and the need for a more comprehensive understanding, we undertook a detailed study of the magnetic properties of Co and Y-doped $CeO_2$ in this paper. The Y ion was chosen because of its non-magnetic cluster nature, which would reveal whether the FM interaction is due to the formation of clusters or oxygen vacancies. To explore the effect of oxygen vacancies further, we annealed the samples in $Ar/H_2$ (a reducing environment) and Air (an oxidizing environment). The reducing nature of the $Ar/H_2$ environment promotes the formation of oxygen vacancies, while the Air environment helps to fill these vacancies. Additionally, we performed Density Functional Theory (DFT) calculations to investigate in detail the effects of dopants and oxygen vacancies on the magnetic properties of $CeO_2$. Although there have been some theoretical studies exploring the impact of Co doping and oxygen vacancies in $CeO_2$, our study provides a more comprehensive analysis, in which we examined the effect of charged oxygen vacancies on magnetic properties and identified the effect of charged oxygen vacancies [23,33,34] Further, we also investigated the most stable charged states of $V_O$. Moreover, we investigated the magnetic coupling between the transition metal ions, adding a crucial layer of understanding to the complex behavior of doped $CeO_2$. This dual approach is believed to provide insights into the correlation between oxygen vacancies-dopant vacancy relation and the observed magnetic behavior. Through these investigations, we aim to resolve some of the existing controversies surrounding the origin of ferromagnetism in $CeO_2$ and contribute to the design of new materials for spintronics and other advanced applications.

## Experimental Procedure

The appropriate stoichiometric amount of Cerium Chloride Heptahydrate [$CeCl_3.7H_2O$; Molecular weight (M.W): 372.58; Made: Sigma Aldrich;99.9%] is used as precursor and Cobalt Acetylacetonate [$Co(C_5H_7O_2)_2$; M.W-257.15; Made: Sigma Aldrich; 97%], Yttrium Chloride Hexahydrate [$YCl_3.6H_2O$, M.W-303.36; Made: Sigma Aldrich; 99.99%] are used as dopant for Co and Y doping respectively. Precursor and dopant solution were prepared separately in triple distilled water with continuous magnetic stirring at 600 rpm. All the reactants were mixed and kept for stirring in a magnetic stirrer for 1 hour to get a homogeneous mixture. Then, ammonium hydroxide ($NH_4OH$) solution was added dropwise to the solution until the pH level reached 10. Further, the solution was kept for stirring for 12 h. Finally, the synthesized precipitate was collected and washed. All precipitates were dried at room temperature and after collection divided into two equal parts. One part is annealed in the furnace in air and another in 9:1 Argon/Hydrogen environments at about 800°C for 8 h. The details of the present co-precipitation method are discussed in our previous work.[35]

After synthesis, the samples were characterized using various analytical techniques. To determine the structural phase of $CeO_2$ and confirm the doping of Co and Y in $CeO_2$, powder X-ray diffraction (XRD) was performed using a Rigaku Smart Lab within the 2θ range of 20–90° with a scan step size of 0.02° using Cu Kα (1.54 Å) radiation. The XRD data were analyzed using FullProf Software (Rietveld method) to assess the cif file for DFT calculations. The morphology and elemental composition of the samples were examined using a Jeol JSM-210 scanning electron microscope (SEM) operated at 20 kV. Energy dispersive spectroscopy (EDS) coupled with SEM was used to verify the incorporation of Co and Y dopants into the $CeO_2$ matrix. X-ray Photoelectron Spectroscopy (XPS, PHI 5000 Versa Probe III) was employed to examine the chemical states of the elements and investigate the effect of different annealing environments on the formation of vacancies. The instrument was calibrated to give a binding energy of 84 eV for Au $4f^{7/2}$ line for metallic gold and all binding energy values were referenced to the C 1s peak at 284.8 eV. To identify the presence of any impurity phases in the samples, room temperature Raman spectra were obtained under air ambient conditions in the 400 $cm^{-1}$ – 800 $cm^{-1}$ range using a Renishaw Raman spectrometer, equipped with a 532 nm Argon ion laser and a confocal microscope. Magnetic measurements were performed using a Superconducting Quantum Interference Device (SQUID) magnetometer (MPMS3, Quantum Design). Magnetization versus field (M-H) loops were recorded over a temperature range of 5 K to 300 K, with the applied magnetic field varying from -6 T to +6 T. Additionally, field-cooled (FC) and zero-field-cooled (ZFC) measurements were performed at 200 Oe in the temperature range of 5 K to 350 K. The optical absorption spectra in the range of 350–800 nm were recorded using a UV-Vis-NIR spectrophotometer (Agilent Cary 5000), which provided insights into the material's optical properties and electronic structure. Photoluminescence (PL) measurements were conducted using a Hitachi F-4600 and a Nanolog spectrometer to study defect-related states within the bandgap. The PL measurements were performed on powder samples that were sonicated in ethanol to ensure uniform dispersion.

**Results and discussion**

After fabricating the $CeO_2$, Co-and Y-$CeO_2$, their surface morphology and elemental analysis to detect the presence of doped elements was studied using SEM (see Supplementary Information, Fig. S1). The impact of TM atoms, specifically Co and Y, as well as the heating environment on the crystal structure of $CeO_2$ were thoroughly examined through XRD measurements. The XRD patterns for all samples are depicted in Fig. 1. The XRD patterns of both pure and doped $CeO_2$ consistently revealed a cubic structure (Fm-3m), aligning with the JCPDS standards having card no: 34-0394, indicative of the polycrystalline nature of the samples. Importantly, the absence of impurity peaks (such as Y, $Y_2O_3$, Co, $Co_2O_3$, $Co_3O_4$, and $CoO_2$) in the spectra reinforces the purity of the investigated samples. However, the observable shift in the doped samples suggests the substitutional doping of Co and Y ions in place of Ce ions within the $CeO_2$ lattice. This shift is reflective of alterations in the lattice constant, as outlined in Table 1. The average lattice constants for all samples were determined using the equation $a^2 = (h^2+k^2+l^2)d_{hkl}^2$ where $d_{hkl}$ is the interplanar distance and (h, k, l) are the respective Miller indices. Specifically, a decrease in the lattice constant was observed for Co-$CeO_2$ Ar/$H_2$, attributed to the smaller ion radius of $Co^{2+}$ (0.065 nm) and $Co^{3+}$ (0.072 nm) compared to $Ce^{4+}$ (0.097 nm) and $Ce^{3+}$ (0.111 nm). This observation aligns with previous reports emphasizing that a smaller dopant ion radius results in a reduced lattice constant.[36] Conversely, an increase in lattice constant was noted for Co-$CeO_2$ Air, indicating lattice expansion, possibly induced by a reduction in the concentration of oxygen vacancies in $CeO_2$. In Y-$CeO_2$ Ar/$H_2$, despite the ion radius of $Y^{3+}$ (0.101 nm) being approximately equal to $Ce^{4+}$ and less than $Ce^{3+}$, a notable lattice compression was observed. This phenomenon could be attributed to a high concentration of $V_O$, facilitated by the Ar/$H_2$ annealing environment. This increases the number of $Ce^{3+}$ ions, contributing to the observed latticecompression.[37] The displacement of oxygen atoms during annealing further contributes to the reduction in the average Ce-O bond length, leading to a decrease in the overall lattice parameter.[38] Meanwhile, Y-$CeO_2$ Air exhibited a similar lattice expansion to Co-$CeO_2$ Air, attributing to a reduction in the concentration of oxygen vacancies compared to Y-$CeO_2$ Ar/$H_2$. Further, the crystallite size (D) is calculated using Debye–Scherrer's formula[39]

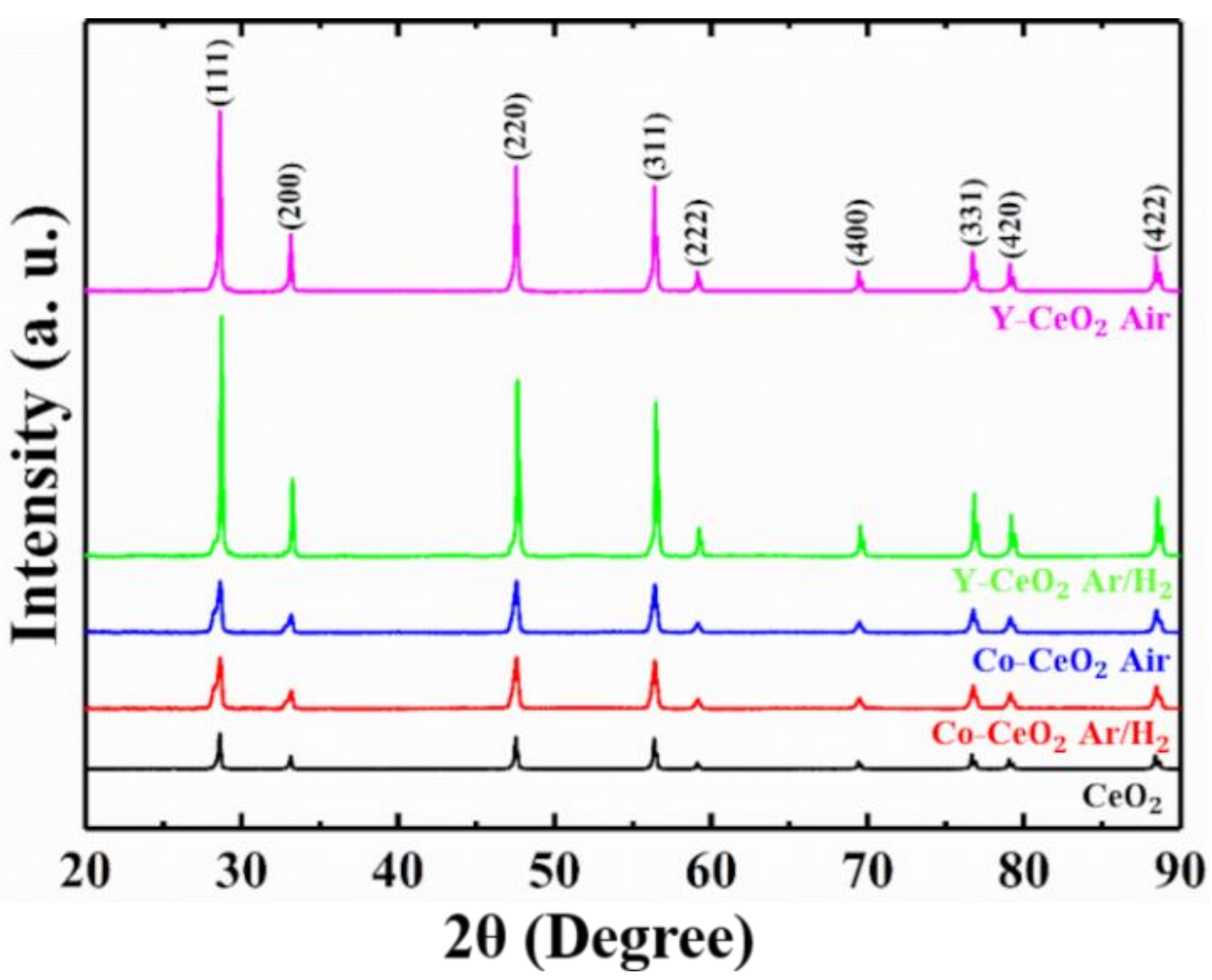

**Figure 1** Room temperature X-ray diffraction patterns of $CeO_2$ and TM (Co and Y) doped $CeO_2$ powders synthesized under Ar/$H_2$ and air atmosphere at 800 ºC for 8 h. The vertical lines are the standard XRD lines of bulk $CeO_2$.

$$D = \frac{0.89\lambda}{\beta cos\theta} \quad (1)$$

Additionally, dislocation density (δ) and strain (ε) are calculated using the equation:

$$\delta = \frac{1}{D^2} \quad (2)$$

$$\varepsilon = \frac{\beta cos\theta}{4} \quad (3)$$

where β is full-width half maximum in radians of all diffraction peaks, θ is the Bragg's angles in degree, and λ is the wavelength of X-ray. The calculated crystallite size, dislocation density, and strain for all samples are listed in Table 1. In this investigation, distinct behaviors were observed for both dopants and the impact of different annealing environments. The study revealed a reduction in the crystallite size of Co-$CeO_2$ Ar/$H_2$ and Co-CeO2 Air, attributed to the small ion radius of Co ions.[40] Conversely, in Y-$CeO_2$ Ar/$H_2$ and Y-CeO2 Air, a slight increase in crystallite size was observed. Similar opposite behaviors were reported in Ni-doped $CeO_2$ thin films.[39] Upon comparing the effects of the heating environment, Co-$CeO_2$ Air exhibited an increase in crystallite size compared to Co-$CeO_2$ Ar/$H_2$, whereas a decrease in crystallite size of Y-$CeO_2$ Air was observed compared to Y-$CeO_2$ Ar/$H_2$.

The calculated δ and ε demonstrated that Co doping in $CeO_2$ led to an increase in dislocation density and strain compared to un-doped $CeO_2$.[41] On the other hand, Y doping resulted in a minimal decrease in dislocation density and strain because of similar ion radius of dopant and host. In Co-$CeO_2$ Air compared to Co-$CeO_2$ Ar/$H_2$ δ and ε values decreased, while in Y-$CeO_2$ Air compared to Y-$CeO_2$ Ar/$H_2$, δ and ε values increased. These findings underscore the distinct impact of the heating environment on different dopants.

**Table 1.** The variation of lattice parameter, crystallite size, dislocation density, and strain in un-doped and TM (Co and Y) doped $CeO_2$ powders.

| | Lattice constant (a) (Å) | Crystallite size (D) (nm) | dislocation density (δ) ($\times 10^{15}$) | Strain (ε) ($\times 10^{-3}$) |
|---|---|---|---|---|
| $CeO_2$ | 5.41360 | 30.2876 | 1.1145 | 1.151 |
| Co-$CeO_2$ Ar/$H_2$ | 5.40903 | 22.7584 | 2.032 | 1.531 |
| Co-$CeO_2$ Air | 5.41838 | 23.8688 | 1.758 | 1.437 |
| Y-$CeO_2$ Ar/$H_2$ | 5.40341 | 32.4932 | 1.049 | 1.093 |
| Y-$CeO_2$ Air | 5.41034 | 31.4593 | 1.099 | 1.122 |

Since defects or spurious phases are challenging to detect using XRD techniques, we have performed more suitable Raman measurements at room temperature. Supplementary Fig. S2 shows the Raman spectra of Co- and Y-doped $CeO_2$ powders annealed in both Ar/$H_2$ and Air environments. The most intense Raman peak at approximately 465 $cm^{-1}$ in all samples corresponds to the symmetrical stretching mode ($F_{2g}$) of the Ce-O8 vibration unit.[42] In the Co-$CeO_2$ Air sample, additional peaks at 522.5, 622, and 691.2 $cm^{-1}$ appear. These peaks are not related to the incorporation of Co into the $CeO_2$ lattice but likely correspond to the presence of $Co_3O_4$ impurities. These peaks are assigned to the $F_{2g}$ (521 $cm^{-1}$), $F_{2g}$ (618 $cm^{-1}$), and $A_{1g}$ (691 $cm^{-1}$) Raman active modes.[43,44] Given their relatively low intensity, it can be inferred that these $Co_3O_4$ grains are very small, on the nanometric scale, which is below the correlation length detectable by X-rays, thus explaining the absence of corresponding peaks in the XRD patterns. In the Co-$CeO_2$ Ar/$H_2$ sample, the intensity of the main Raman peak is reduced, which could be attributed to changes in the crystallite size. However, XRD analysis indicates that the crystallite size remains approximately the same. The reduction in Raman intensity is likely due to an increase in dislocation density and strain in the Co-$CeO_2$ Ar/$H_2$ sample. Conversely, the Y-$CeO_2$ Ar/$H_2$ sample shows a higher intensity of the main peak, indicating a decrease in strain and dislocation density, which is consistent with the findings from the XRD analysis.

The oxidation state of Ce in $CeO_2$ and the creation of oxygen vacancies due to varying environments are key factors, elucidated by the presented XPS analysis. The analysis of XPS survey spectra indicates the presence of two major components, Ce and O, on the surface, along with a substantial amount of Co and Y doping as shown in Supplementary Fig. S3. The deconvoluted Ce 3d spectra in Supplementary Fig. S4 provide insights into electron transitions between $Ce^{4+}$ and $Ce^{3+}$ states. $Ce^{4+}$ states are identified by emissions from Ce $3d_{5/2}$ and Ce $3d_{3/2}$ core levels, denoted as v, v'', v''', u, u'', and u'''. The $Ce^{3+}$ state involves different final states, represented by $v_0$, v', $u_0$, and u' transitions. Based on the previous literature, the binding energy of the major components v (~879-881 eV), u (~898-900 eV), v''' (~888-890 eV), u''' (~914-916 eV), and the minor components v'' (~885-888 eV) and u'' (~905-907 eV), is due to $Ce^{4+}$ cations. Whereas, for $Ce^{3+}$ cations, the v' (~881-885 eV) and u' (~900-904 eV) components are found to be more intense. On the other hand, the $v_0$ (~876–879 eV) and $u_0$ (~896–898 eV) components are very less prominent. Their signals in the Ce 3d spectrum are often diminished particularly when there is a high concentration of $Ce^{4+}$ in the samples.[45]

The deconvoluted O 1s spectra as shown in Fig. 2 exhibit distinctive features across various samples. Typically, the O 1s core levels at approximately ~528-529 eV ($O_L$) correspond to $O^{2-}$ ions within the cubic structure.[46] The binding energy at around ~530-531 eV ($V_O$) indicates the non-lattice oxygen/ surface adsorbed OH group.[47] Peaks $O_\alpha$ observed at approximately ~532-533 eV are associated with weakly adsorbed species, such as ether groups. Peaks $O_\beta$, detected at around ~534-535 eV are associated with single-bonded oxygen present in carboxylic acid or ester groups (or chemisorbed water) adsorbed on the $CeO_2$ surface.[48–50] Notably, the Ar/$H_2$ annealed sample exhibits a higher concentration of lattice oxygen and non-lattice oxygen/OH compared to the air-annealed sample, as listed in Table 2. The increase in the area of non-lattice oxygen indicates the formation of oxygen vacancies in the lattice, highlighting the significant influence of the annealing environment on the concentration of oxygen vacancies in the sample.[47] Furthermore, a detailed analysis of surface-adsorbed carbon groups is provided through the deconvoluted C 1s spectra shown in Supplementary Fig. S5. This analysis reveals the presence of various carbon-containing groups in the air-annealed sample, whereas these groups are either absent or present in significantly lower amounts in the Ar/$H_2$ annealed samples.

**Table 2.** Fitted peak area (%) for lattice oxygen ($O_L$), oxygen vacancies ($V_O$), and Ce 3d oxidation states ($Ce^{3+}$ and $Ce^{4+}$).

| | $O_L$ | $V_O$ | $Ce^{3+}$ | $Ce^{4+}$ |
|---|---|---|---|---|
| $CeO_2$ | 41.50 | 33.79 | 29.27 | 52 |
| Co-$CeO_2$ Ar/$H_2$ | 66.65 | 33.34 | 31.21 | 62.48 |
| Co-$CeO_2$ Air | 32.09 | 20.24 | 30.94 | 59.69 |
| Y-$CeO_2$ Ar/$H_2$ | 54.29 | 37.16 | 35.08 | 52.6 |

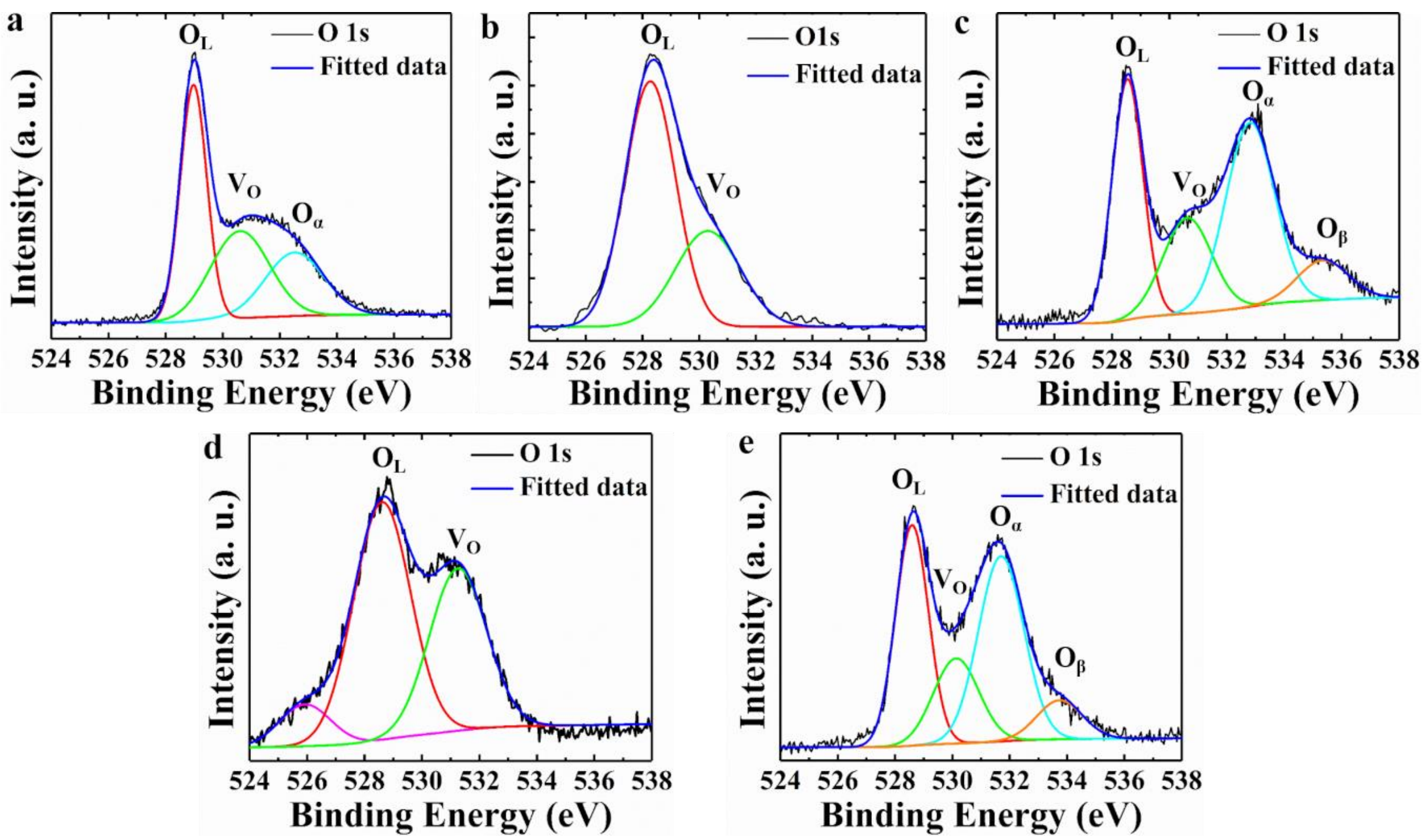

**Figure 2** O 1s spectra of (a) $CeO_2$, (b) Co-$CeO_2$ Ar/$H_2$, (c) Co-$CeO_2$ Air (d) Y-$CeO_2$ Ar/$H_2$, and (e) Y- $CeO_2$ Air.

| Y-$CeO_2$ Air | 34.11 | 18.14 | 30.83 | 62.18 |
|---|---|---|---|---|

The XPS spectrum of Co 2p in Supplementary Fig. S6(a, b) indicates the presence of both Co $2p_{3/2}$ and Co $2p_{1/2}$ peaks, confirming the existence of Co ions in the sample. In the Co-$CeO_2$ Ar/$H_2$ sample, the observed core-level binding energies of the Co $2p_{3/2}$ and Co $2p_{1/2}$ peaks are 777.18 eV and 792.68 eV, respectively, with an energy separation of approximately 15.5 eV.[51,52] These core-level binding energies are consistent with the presence of $Co^{2+}$ ions. The satellite peaks further reveal a high-spin Co(II) state with complex transitions. The position of the core-level peaks at lower binding energies may be attributed to the presence of vacancies or defects, as evidenced by the O 1s XPS spectra and PL measurements described below. These results indicate that Co does not precipitate as metallic Co on the surface. In the Co-$CeO_2$ Air sample, the peaks at 779.5 eV and 782.2 eV were assigned to Co $2p_{3/2}$, while those at 794.3 eV and 797.7 eV correspond to Co $2p_{1/2}$, indicating the presence of both $Co^{3+}$ and $Co^{2+}$ oxidation states. The concentration of $Co^{3+}$ peaks, accounting for approximately 43%, is accompanied by some impurity phase of $Co_3O_4$, as supported by the Raman analysis. The XPS spectra of Y 3d in Supplementary Fig. S6(c, d) exhibit two prominent sub-peaks at $3d_{5/2}$ and $3d_{3/2}$ due to the high and low binding states of Y $3d_{3/2}$ and Y $3d_{5/2}$, respectively, confirming the presence of $Y^{3+}$.[53] Interestingly, in the Y 3d spectra, the intensity of $3d_{5/2}$ is higher than that of $3d_{3/2}$ in Y- $CeO_2$ Ar/$H_2$, while the opposite behavior is observed in Y- CeO2 Air. This variation in intensity between $3d_{5/2}$ and $3d_{3/2}$ peaks could be influenced by factors such as the local binding environment, neighboring atoms, and chemical bonding.

It's important to note that the relative intensities of XPS peaks are sensitive to the specific chemical state, electronic configuration, and local environment of the elements being analyzed. The observed differences in peak intensities provide valuable insights into the chemical and electronic changes induced by different annealing environments and highlight the role of atmospheric conditions in influencing the surface chemistry of the samples.

The optical properties of both un-doped and TM-doped $CeO_2$ samples are shown in Fig. 3. A distinct absorption peak around 360 eV is observed in all samples, indicating a charge-transfer transition from O 2p to Ce 4f state in $CeO_2$.[24] The optical bandgap ($E_g$) of all samples is determined using the formula[54]

$$(\alpha h\nu)^{1/n} = A(h\nu - E_g) \quad (4)$$

where α is the absorption coefficient, hν represents the energy of the photons, h is Planck's constant, ν is the vibration frequency, and A is a proportionality constant. The exponent n depends on the nature of the semiconductor's optical transition, where n = ½ for direct transitions and n = 2 for indirect transitions. Further, the $E_g$ of the nanoparticle is determined by extrapolating the straight line portion of the plot $(\alpha h\nu)^2$ verses hν. The obtained $E_g$ values are listed in the inset of Fig. 3(b). The obtained $E_g$ value in this experiment is slightly less compared to the reported values for $CeO_2$ particles.[55] This discrepancy may be attributed to the presence of defect states, as suggested by the XPS studies conducted earlier.[56,24] The introduction of defect states alters the electronic structure of both un-doped and doped $CeO_2$, resulting in a narrower band gap and a red shift in the absorption band edge.[18] In Co-$CeO_2$, the bandgap value is significantly lower compared to other

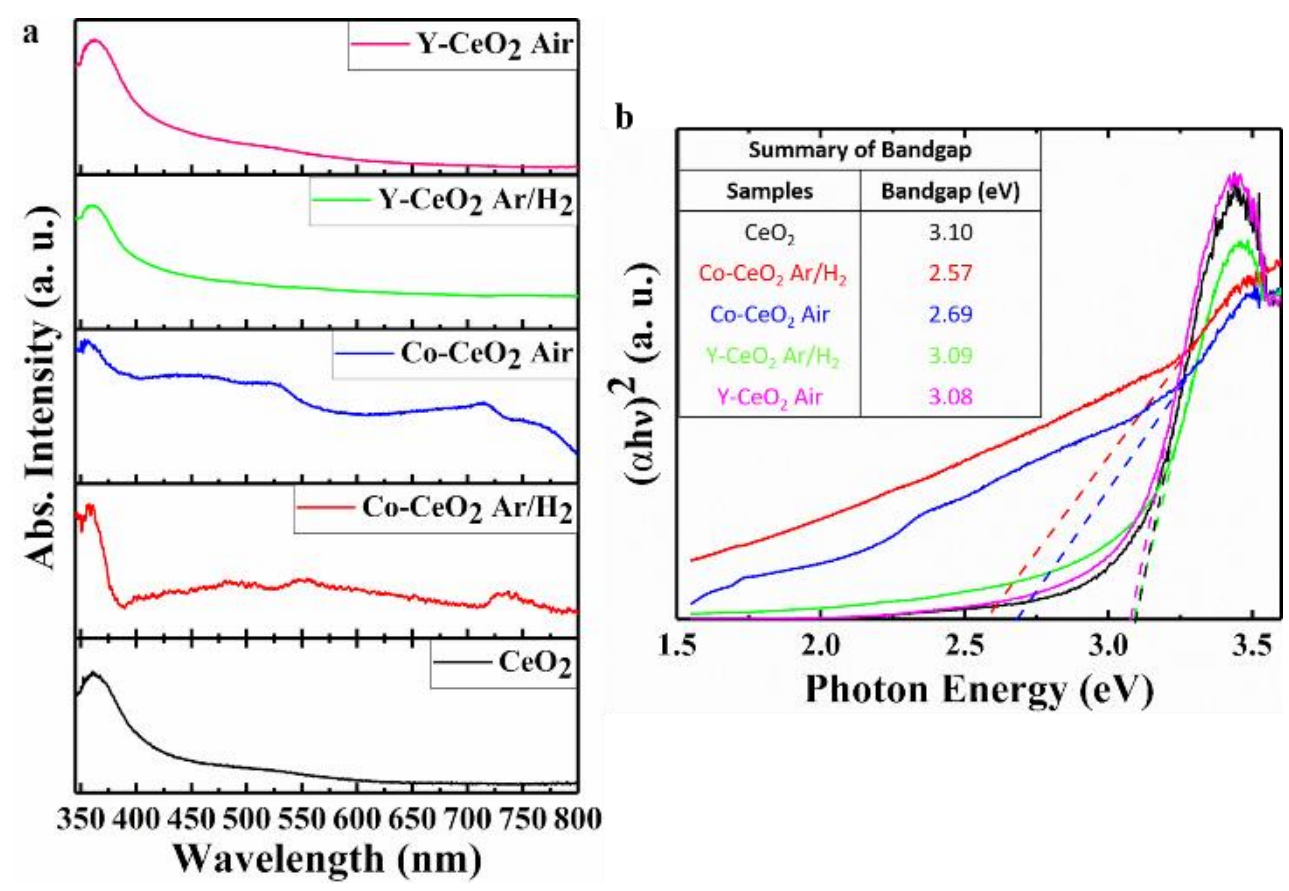

**Figure 3** (a) UV–Vis. diffuse reflectance of (a) Absorption spectra of all samples and (b) Energy band gap of $CeO_2$ and TM–doped $CeO_2$ powder using Tauc plot with an inset showing the band gap values of samples.

samples, indicating that Co ions induce more defect states, which facilitate the reduction in the band-gap energies.[57] These Co ions can create mid-gap states that act as electron trap sites or facilitate intermediate electronic transitions. As a result, the material may exhibit continuous absorption in the UV-Vis range. In both Y-$CeO_2$ Ar/$H_2$ and Y-$CeO_2$ Air samples, the band gap is slightly less compared to $CeO_2$, indicating the presence of $Ce^{3+}$ states near the conduction band due to the formation of oxygen defects. The similar absorption behavior observed in both un-doped and Y-$CeO_2$ suggests that the Y ion does not introduce significant defect states within the band gap that would lead to continuous absorption beyond the band gap. This behavior is consistent with the results obtained from the DFT calculations, as discussed below.

PL spectroscopy was also employed to investigate the optical properties and defect concentrations in the samples, as depicted in Supplementary Fig. S7. By using Gaussian fitting, approximately seven peaks were observed, as illustrated in the spectra. The peaks present in the 360–400 nm range (corresponding to 3.45–3.10 eV) are attributed to the band-edge emission of the samples, indicative of charge transfer from the Ce 4f to O 2p level. Moreover, the peaks in the 400–550 nm range (3.10–2.25 eV) may be related to F-centers, including $F^{2+}$ of the oxygen vacancy having no trapped electron, $F^{+}$ with oxygen vacancy having single trapped electron, and $F^{0}$ with oxygen vacancy having two trapped electrons centers.[58] The presence of $Ce^{3+}$ ions and oxygen defects was confirmed through XPS analyses. Notably, the PL intensity of the Co-$CeO_2$ Air sample was observed to be lower than that of the Ar/$H_2$ annealed sample, indicating defects predominantly manifest as surface defects, serving as trap centers. Consequently, these trap centers act as non-radiative recombination centers, thereby suppressing the emission intensity.[59] Whereas, the PL intensity of Y-$CeO_2$ Ar/$H_2$ and Air samples are approximately similar, which signifies that the samples have similar electronic and optical properties.

The magnetization characteristics were examined by measuring variations with applied magnetic fields for both un-doped and doped samples at temperatures ranging from 5 K to 300 K, as depicted in Fig. 4. The M-H loops of all samples illustrate that only Ar/$H_2$ annealed Co- and Y-$CeO_2$ exhibit saturation magnetization at low fields, suggesting that single- and multi-domain samples account for the magnetization. However, even at the strongest field and lowest temperature, the air-annealed samples exhibit non-saturation in magnetization, confirming a paramagnetic-dominated

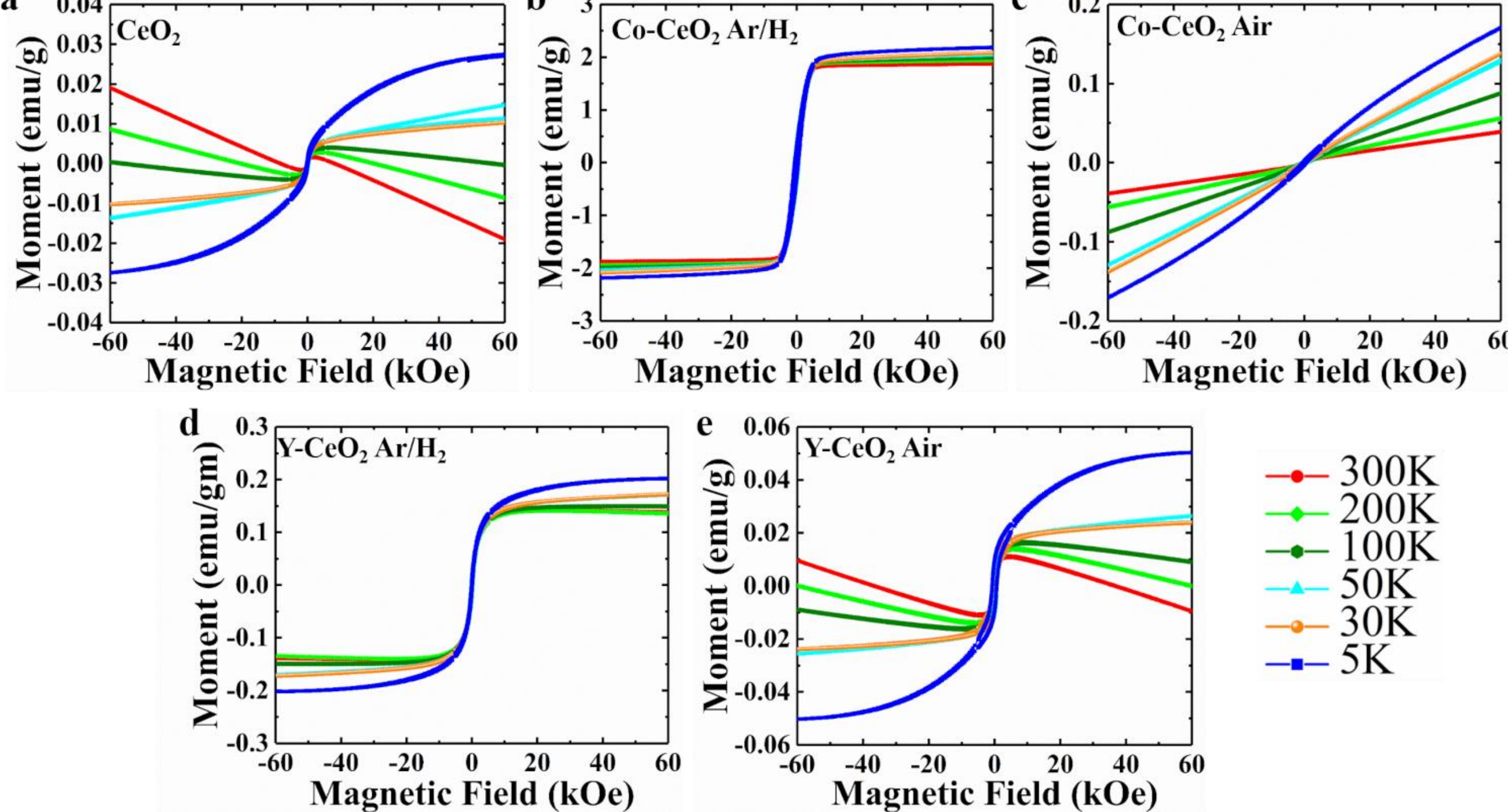

**Figure 4** Magnetization hysteresis loops (a) $CeO_2$ (b) Co-$CeO_2$ Ar/$H_2$, (c) Co-$CeO_2$ Air, (d) Y-$CeO_2$ Ar/H2, (e) Y-$CeO_2$ Air.

magnetic contribution. Nevertheless, at higher temperatures, the magnetization of air-annealed Y-doped $CeO_2$ samples diminishes as the magnetic field increases, indicating a significant diamagnetic response in this specimen. Hence, the Co-$CeO_2$ Air sample comprises two magnetic phases dominantly paramagnetic and weakly FM, whereas the Y-$CeO_2$ Air is composed of all three phases (weakly FM, strongly paramagnetic, and diamagnetic). From an application standpoint, the ferromagnetically dominated Ar/$H_2$ annealed samples are superior to air-annealed samples. To validate this fact quantitatively, only the FM contribution after subtraction of the paramagnetic and diamagnetic contributions is depicted

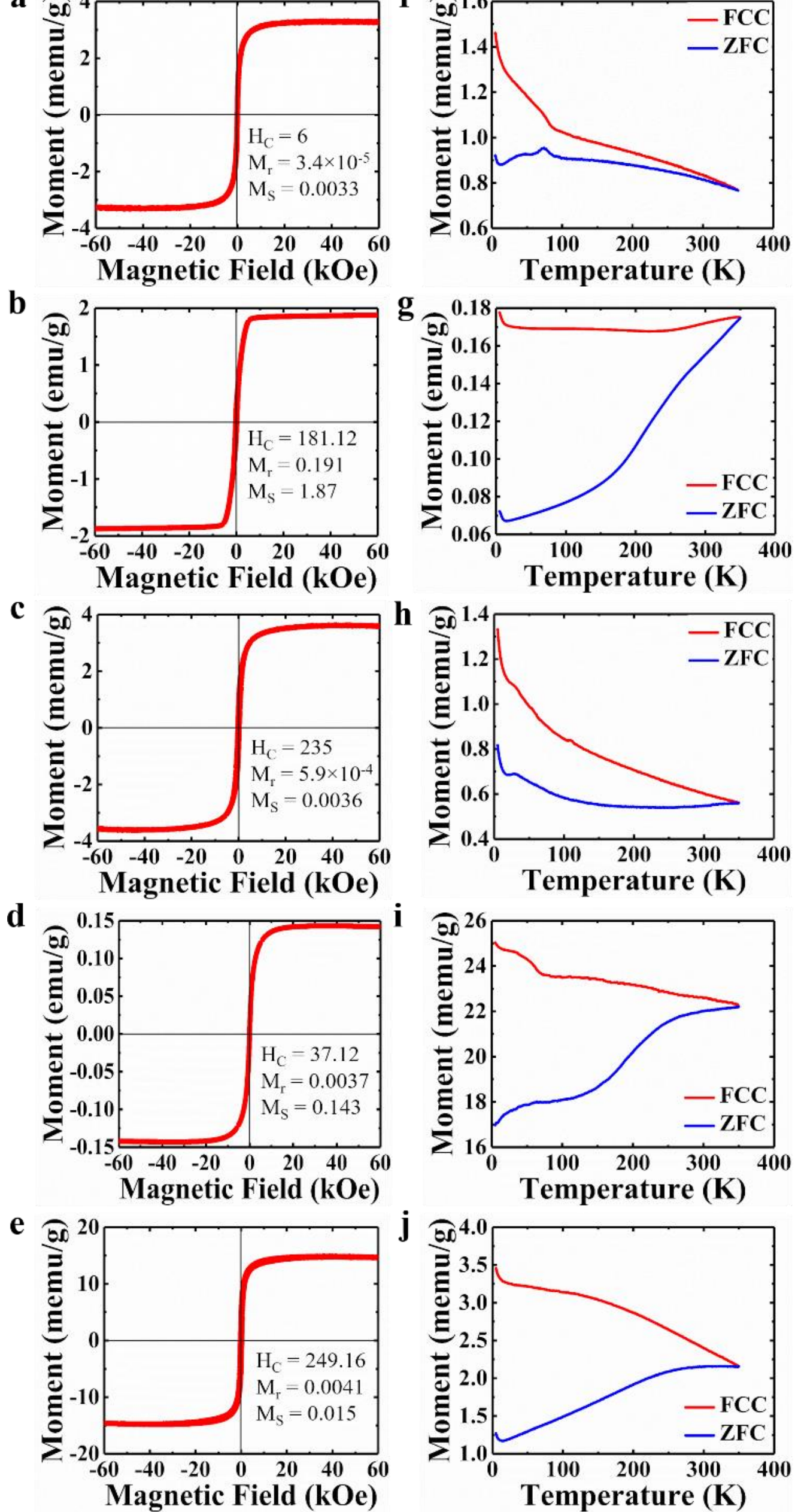

**Figure 5** (a-e) M–H hysteresis curves at room temperature after subtraction of paramagnetic or diamagnetic contribution, (f-j) ZFC and FC variation of the magnetization as a function of temperature of $CeO_2$, Co-$CeO_2$ Ar/$H_2$, Co-$CeO_2$ Air, Y-$CeO_2$ Ar/$H_2$, and Y-$CeO_2$ Air.

in Fig. 5(a-e). The coercive field ($H_C$), remanent magnetization ($M_r$), saturation magnetization ($M_S$), and squareness ratio ($S = M_r/M_S$) values obtained from the hysteresis loop at 300 K are listed in the inset of Fig. 5(a-e). The $H_c$ of un-doped $CeO_2$ is measured at 6 Oe, which increases to 181 Oe and 37 Oe for Co- and Y-$CeO_2$ Ar/$H_2$, respectively. Further increases in coercivity values are observed for Co- and Y-doped $CeO_2$ Air, reaching 235 Oe and 249 Oe, respectively. Coercivity is indicative of the magnetic field strength required to overcome magnetic anisotropy. In air-annealed samples, the magnetic anisotropy is higher compared to Ar/$H_2$ annealed samples. The high $H_c$ in air-annealed samples is due to the suppression of oxygen vacancies, where the structural defects in an air environment lead to a more ordered crystalline lattice with fewer defects. It is clear that upon annealing in an Ar/$H_2$ environment, there is a noticeable increase in the $M_S$ value. Note that the $M_S$ of air-annealed samples is in the mili-emu range, while that of Ar/$H_2$ annealed samples is in the emu range. This observation substantiates that the Ar/$H_2$ annealed sample contains a single magnetic phase. The Co- and Y-$CeO_2$ Ar/$H_2$ samples have a far higher magnetization than recently published doped $CeO_2$.[60–64] Based on these observations, we can conclude that the Co-$CeO_2$ Air annealed sample exhibits characteristics of magnetically clean samples due to its quite low value of $M_r$. The Y-$CeO_2$ Air sample demonstrates a high $H_c$ and a low $M_r$ value, but the presence of multiple magnetic phases hinders the possibility of its application in spintronics.

Further, to understand the magnetic behavior and magnetic transition temperatures we have performed temperature-dependent Zero Field Cooled (ZFC) and Field Cooled (FC) magnetization measurements for all samples at 200 Oe, which are presented in Fig. 5(f-j). The nature of ZFC and FC curves vary among samples, with bifurcation observed in all samples. The temperature at which the ZFC and FC curves bifurcate is known as the bifurcation temperature ($T_{bif}$). This bifurcation is commonly attributed to the largest particle size distribution.[65] In our samples, $T_{bif}$ is approximately greater than 350 K and the samples exhibit above-RTFM behavior. In the ZFC curves of all samples, no well-defined transition temperature peak is observed which corresponds to a wide range distribution of particle size. In un-doped $CeO_2$, The ZFC and FC curves indicate mixed paramagnetic and FM behavior. Upon decreasing the temperature, both ZFC and FC increase, suggesting paramagnetic behavior. In the ZFC and FC curves, a transition is observed around 70 K, where ZFC starts decreasing, and a peak is found.[66] In Co-$CeO_2$ Air, both the ZFC and FC curves show a continuous increase in magnetization as the temperature decreases, which is characteristic of paramagnetic behavior. A small kink around 30 K was observed in the magnetic measurements, consistent with the antiFM (AFM) transition which indicates the presence of the $Co_3O_4$ impurity. This AFM phase is likely responsible for the diminished FM behavior in the Co-$CeO_2$ Air sample.

The small difference between ZFC and FC moments indicates less FM behavior. Based on the above behavior M-T curve (FC

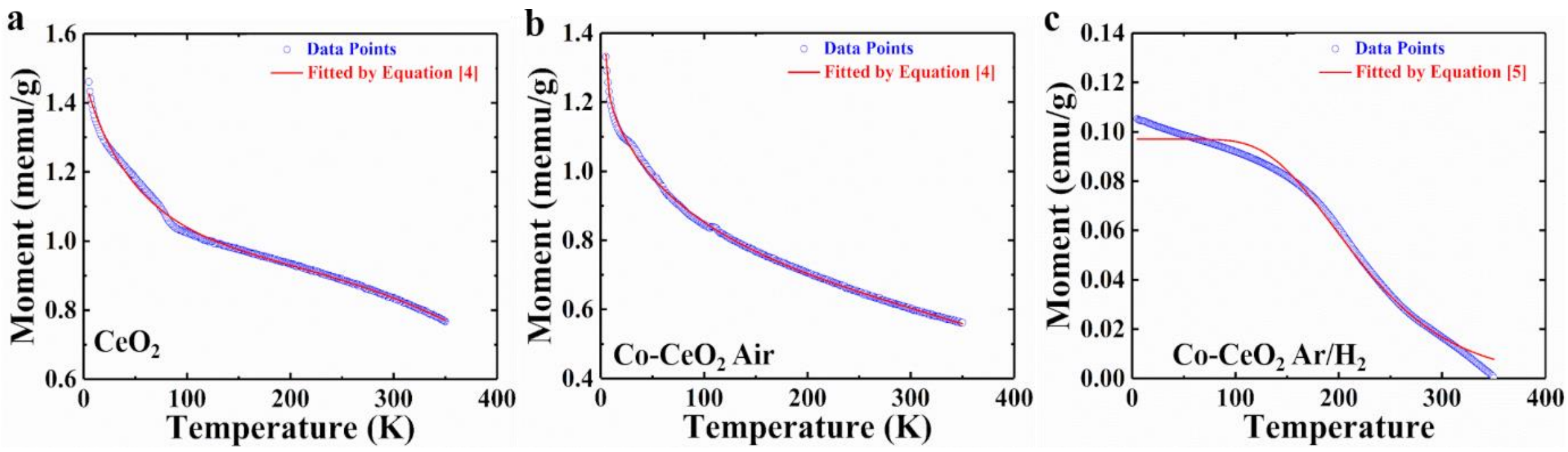

**Figure 6** (a) The Fitting of FCC MT curve of (a) $CeO_2$, (b) Co-$CeO_2$ Air, and (c) The fitting of the difference between the FC and ZFC magnetizations ΔM as a function of temperature T curve of Co-$CeO_2$ Ar/$H_2$.

mode) of un-doped $CeO_2$ and Co-$CeO_2$ Air is fitted by modified Bloch's law[67]

$$M(T) = M(0)\left[1 - BT^{b} - FT\ln T - CT\right] + \frac{M}{(T-\theta)} \qquad (5)$$

Further, b is taken as 1.5 according to mean field theory, and fitting yields B = 2.17 $\times 10^{-4}$ and 1.96 $\times$ $10^{-4}$, which is an indication of FM NPs. B's positive and low values signify a positive exchange interaction between the spins. In the case of the bulk system, the value B is the order of $\sim 10^{-6}$, almost 100 times increase is indicative of small NC formation.[67] The M-T curves are well-fitted for un-doped and air annealed samples as shown in Fig. 6(a, b). The good fitting substantiates the presence of weak FM along with PM.

In Co-$CeO_2$ Ar/$H_2$, a decrease in the freezing of the moment in the FC and ZFC curve is observed without showing any peak suggesting that below RT the magnetic moments of the system become locked or "frozen" in the direction of the applied magnetic field. This behavior is characteristic of a spin glass/superparamagnetic system. To check further whether all particles are single domain, have tried to fit the difference between the FC and ZFC magnetization ΔM ($M_{FC}$-$M_{ZFC}$) by the equation mentioned below for a pure superparamagnetic (SPM) system.[68,69]

$$\Delta M = \frac{29 M_S^2 H}{6K_a}\left\{1 - \mathrm{erf}\left[\frac{\ln(T/T_{B0})}{\sqrt{2}\, c} - \frac{\lambda_B}{\sqrt{2}}\right]\right\} \qquad (6)$$

Here, we consider the mean blocking temperature

$$< T_B >= T_{B0}\, exp\left(\frac{{\lambda_B}^2}{2}\right) \qquad (7)$$

The error function is defined as

$$\int_T^{\infty} e^{-t^2} dt = \left(\sqrt{\pi}/2\right) erfc(T), \qquad (8)$$

where, $erfc(T) = 1 - erf(T)$

The poor fitting of the magnetization data by the above equation established that none of the samples are purely SPM (see Fig. 6(c)). This may be due to the uncompensated surface spin fluctuations, which are responsible for short-range magnetic correlations. These correlations may form exchange-coupled clusters and yield collective freezing (frustrated magnetic state).[70] Another possibility is that the magnetic interactions are strong enough to suppress a clear peak, potentially due to the clustering of magnetic ions or strong exchange interactions. Extending the Y doping, it is observed that in Y-$CeO_2$ Ar/$H_2$, the separation between FC and ZFC is higher compared to Y-$CeO_2$ Air, indicating a higher FM characteristic. The low moment of Y-$CeO_2$ is also an indication that the presence of oxygen vacancy as well as the type of dopant both are important for ensuring high magnetization. In all samples except Y-$CeO_2$ Ar/$H_2$, a sudden increment at very low temperatures below 12 K, is observed, possibly originating from loose spins/non-correlated spin on the surface of the particles. [71]

The origin of ferromagnetism in $CeO_2$ has been well studied and is primarily attributed to the formation of oxygen vacancies ($V_O$).[72] The FM interaction is enhanced by the electrons released from these oxygen vacancies, which may be trapped by $Ce^{4+}$ ions, reducing them to the $Ce^{3+}$ state and leading to the formation of ionized vacancy states. The exchange interaction between the electron in the ionized vacancy state and the electron in the Ce ions favors FM ordering.[73] In TM-doped $CeO_2$, the observed rise in ferromagnetism can be attributed to several factors, including the intrinsic properties of the doped NPs, the F-center exchange (FCE) mechanism, the formation of TM clusters or impurity phases, and superexchange or double exchange magnetic coupling between the TM ions. While several studies have suggested that the formation of TM clusters contributes to the FM interaction in DMOs, this does not apply to our case. Specifically, FM behavior is observed in both Co- and Y-$CeO_2$ Ar/$H_2$ samples. However, TM clusters typically induce ferromagnetism only when the TM ions themselves exhibit magnetic behavior. If the FM behavior in Co-$CeO_2$ were due to the formation of Co clusters, a similar mechanism should be observed in Y-doped $CeO_2$, as no other impurity phases were observed. However, it is well known that $Y^{3+}$ ions are non-magnetic. Therefore, the formation of FM clusters is not possible, ruling out TM cluster formation as the mechanism behind the FM interaction in these samples. If Co clusters were present, we could estimate their contribution to the magnetic moment based on the magnetic moment of metallic Co, which is 1.72 $\mu_B$ per Co atom. For the observed magnetic moment of 1.12 $\mu_B$ per Co atom, approximately 65% of the doped Co would need to form clusters.[23] This large fraction of Co in cluster form would likely be detectable in the XPS 2p spectrum of Co. However, no metallic Co is observed in the XPS data, further supporting the conclusion that TM cluster formation is not responsible for the observed ferromagnetism. Furthermore, no impurity phase responsible for FM coupling was detected from the characterization, the only identified impurity phase was $Co_3O_4$, which was observed exclusively in the Co-$CeO_2$ Air sample. $Co_3O_4$ exhibits AFM coupling, which further supports that the observed FM behavior cannot be attributed to impurity phases. In both Co- and Y-$CeO_2$ Ar/$H_2$ samples, Co is in the $Co^{2+}$ oxidation state and Y in the $Y^{3+}$ oxidation state, which rules out double exchange interactions between these ions. In the Co-$CeO_2$ system, FM interactions between $Co^{2+}$ ions, such as $Co^{2+}$-$O^{2-}$-$Co^{2+}$, $Co^{2+}$-$V_O$-$Co^{2+}$, and $Co^{2+}$-$V_O$-$Ce^{3+}$ are possible.[36] However, such interactions do not occur in Y-doped samples, since $Y^{3+}$ is non-magnetic. The primary mechanism responsible for the observed FM behavior in TM-doped $CeO_2$ is the FCE mechanism, which has been well-documented in the literature and is consistent with our results. In our study, the Ar/$H_2$ annealed samples show a higher concentration of oxygen vacancies and exhibit stronger FM interactions. This correlation clearly suggests that the FCE mechanism plays a dominant role in the ferromagnetism observed in our samples.

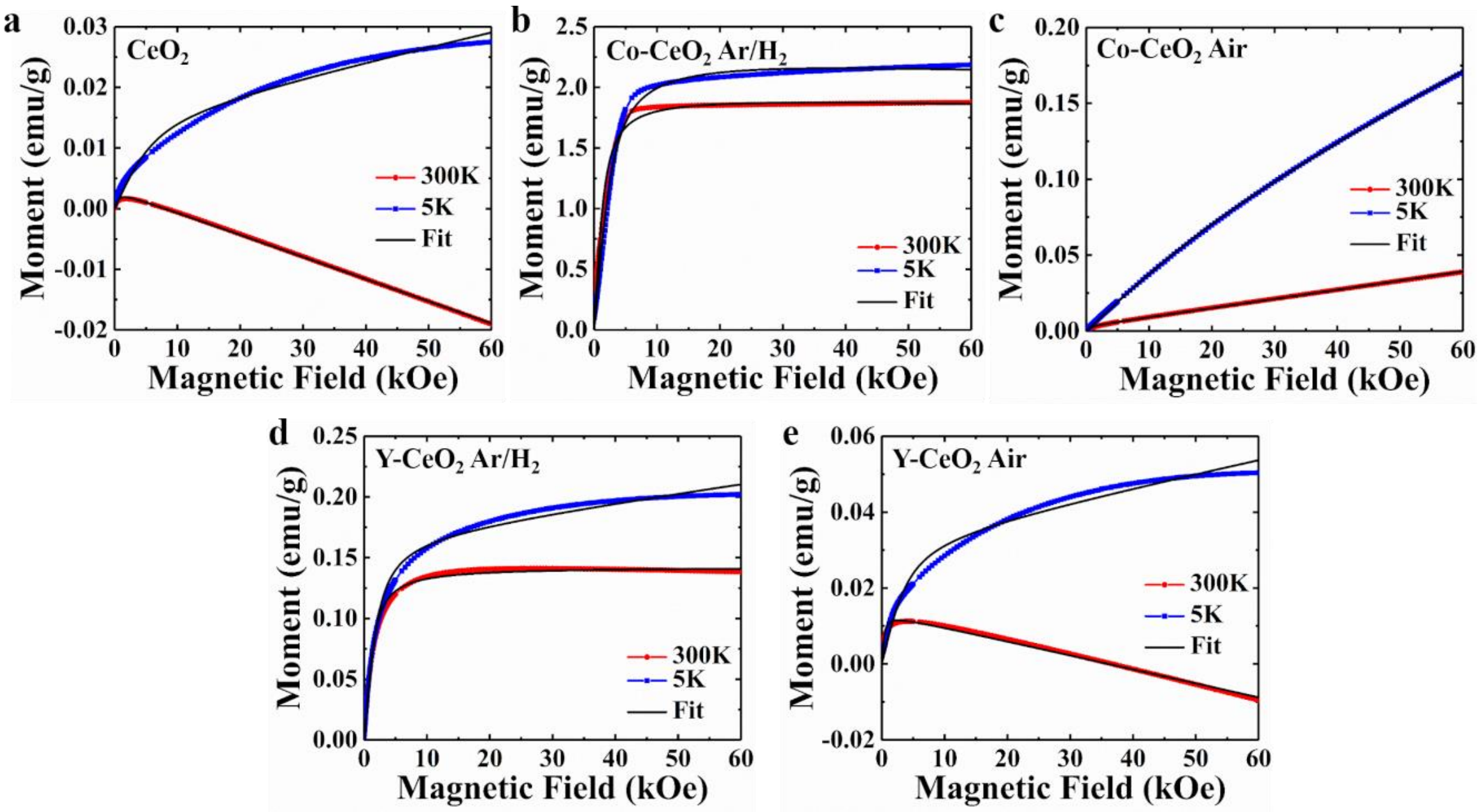

**Figure 7** Magnetization hysteresis loops (a) $CeO_2$ (b) Co-$CeO_2$ Ar/$H_2$, (c) Co-$CeO_2$ Air, (d) Y- $CeO_2$ Ar/$H_2$, (e) Y- $CeO_2$ Air.

All the above discussion concludes that the FM in doped $CeO_2$ arises from the exchange interactions between dopant ions mediated by vacancies known as the F center exchange mechanism. The observed FM is likely attributed to the existence of surface $V_O$, which can lead to the formation of $F^+$ centers.[74] The $F^{2+}$ and $F^0$ vacancies contribute to the decrement in long-range FM interaction in all samples.[23,75] Furthermore, the BMP model attributes the origin of RTFM in DMOs to the presence of defects in the sample. BMPs, regarded as isolated FM entities, form orbits with finite diameters similar to hydrogenic Bohr orbits due to electron localization in defect centers.[76] The characteristics of the M-H curves can also be elucidated using the BMP model. The unsaturated saturation of the curves indicates that BMP concentrations are lower than the percolation threshold needed for bulk ferromagnetism. In the core of the nanocrystals, BMPs may not form because of the deficiency of $V_O$. A substantial net spontaneous magnetic moment is possessed by each BMP, which aligns swiftly in response to a magnetic field. Consequently, magnetization is suddenly increased even at weak magnetic fields. The spins of the host or dopant outside the BMP align comparatively slowly along the applied magnetic field.[77] Consequently, it is only at the higher magnetic fields that the contribution from the diamagnetic and paramagnetic aligned core becomes predominant, resulting in a diamagnetic and paramagnetic tail.

To separate the impact of BMPs from those of the paramagnetic and diamagnetic matrix, the above-discussed BMP Model was employed. Firstly, the true spontaneous moment ($m_s$) of a single BMP was determined and the number of BMPs per $cm^3$ (N) was calculated. In the BMP model, the FM nature was adequately described using the Langevin profile equation[78]:

$$L(x) = \coth(x) - \left(\frac{1}{x}\right) \quad (9)$$

where $x = m_{eff}H/K_BT$; $m_{eff}$ represents the effective magnetic moment per BMP and $K_B$ denotes the Boltzmann's constant. The FM component of the equation, denoted as, $M_{FM}$, can be expressed using saturation magnetization $M_0$, as $M_{FM} = M_0L(x)$ ($M_0$ is the product of N and $m_s$). The alignment rate of the true moment along H is governed by the effective moment ($m_{eff}$) in the Langevin function's argument. In the collective regime, there is no distinction between $m_{eff}$ and $m_s$. Furthermore, in the high-field region, the presence of the diamagnetic and paramagnetic components can be denoted in terms of the susceptibility $\chi_m$. If $\chi_m < 0$, it represents diamagnetic behavior, while if $\chi_m > 0$, it indicates paramagnetic behavior. Thus, the total magnetization data were fitted using the Langevin linear equation from the M–H curves:

$$M = M_{FM} + M_{PM} \quad (10)$$

$$M = M_0 L(x) + \chi_m H \quad (11)$$

where $M_0$, $m_{eff}$, and $\chi_m$ denote the fitting parameters. Fitting of the BMP model to un-doped and doped samples was conducted to obtain these parameters. The values of these parameters obtained from the fitting of the 300 K and 5 K M-H curves are summarized in Table 3 and Table 4, respectively. The fitted data demonstrated a close agreement with the experimental evidence as illustrated in Fig. 7(a-e). The BMP concentration is maximum in $Ar/H_2$ annealed samples, leading to the highest FM contribution observed in Co and Y-doped $CeO_2$ $Ar/H_2$ samples.

As evident from Table 3, the number of BMPs varies with changes in the type of TM ions and the annealing environment. BMPs are formed by the localized charge carriers trapped by $F^+$ defect centers and the presence of numerous surrounding TM ions by aligning all the TM spins around the carrier localization center. Long-range magnetic interaction in the system can occur either through direct overlap of two polarons or from indirect interactions mediated by a magnetic impurity between two polarons. The behavior of these interactions finally depends on the number density of the formed BMPs and their separation within the host matrix. The BMP number density in both un-doped and TM-doped $CeO_2$ samples has been estimated to be around $10^{15}$ $cm^{-3}$ at room temperature and increases to about $10^{17}$ $cm^{-3}$ at 5 K. At low temperatures, electron spin has the ability to align the localized spins, leading to an increase in the number of BMPs. However, this density is not high enough for direct overlap and hence reaches the limit of percolation. Therefore, the observed lack of $M_0$ can be attributed to this limitation. The exchange interaction is enhanced among the BMPs due to the magnetic impurity serving as the mediator to achieve ferromagnetism in the system.[78] The paramagnetic and diamagnetic contributions stem from F-centers and the $4f^0$ electronic configuration of $Ce^{4+}$ ions, respectively.

The increased FM behavior observed in Co and Y-doped $CeO_2$ $Ar/H_2$ samples indicates a higher concentration of BMPs. Additionally, in the Co-$CeO_2$ $Ar/H_2$ sample, there is higher magnetization than in Y-$CeO_2$ $Ar/H_2$ because Co ions spin around the carrier localization center, whereas in the Y-$CeO_2$ $Ar/H_2$ sample, Y ions have non-magnetic nature, so magnetism primarily originates from the contribution of $Ce^{3+}$ ions located at the surface of the sample. Conversely, the suppression of BMPs in air-annealed samples, despite the presence of vacancies and magnetic impurities (predicted by XPS and PL), leads to diminished FM behavior. This decrease can be attributed to the presence of $F^{2+}$ and $F^0$ vacancies, which contribute to a decrease in long-range FM interaction across all samples.

**Table 3.** Experimental and fitting data obtained from the BMP model of M–H curves for un-doped and doped $CeO_2$ at 300 K.

| Samples | $M_S$ (emu/g) | $M_S$ ($u_B$/TM ion) | $M_0$ (emu/g) | $X_m \times 10^{-6}$ ($u_B$/TM ion) | $M_{eff} \times 10^{-17}$ (emu) | $N \times 10^{15}$ ($cm^3$) |
|---|---|---|---|---|---|---|
| $CeO_2$ | 0.0033 | - | 0.003 | -0.366 | 13.66 | 0.022 |
| Co-$CeO_2$ $Ar/H_2$ | 1.87 | 1.12 | 1.95 | -1.076 | 5.88 | 33.16 |
| Co-$CeO_2$ Air | 0.0036 | 0.0024 | 0.003 | 0.601 | 11.96 | 0.025 |
| Y-$CeO_2$ $Ar/H_2$ | 0.143 | 0.083 | 0.143 | -0.00129 | 5.84 | 2.45 |
| Y-$CeO_2$ Air | 0.015 | 0.007 | 0.0134 | -0.369 | 22.11 | 0.061 |

**Table 4.** Experimental and fitting data obtained from the BMP model of M–H curves for un-doped and doped $CeO_2$ samples at 5 K.

| Samples | $M_S$ (emu/g) | $M_S$ ($u_B$/TM ion) | $M_0$ (emu/g) | $X_m \times 10^{-6}$ ($u_B$/TM ion) | $M_{eff} \times 10^{-19}$ (emu) | $N \times 10^{17}$ ($cm^3$) |
|---|---|---|---|---|---|---|
| $CeO_2$ | 0.027 | - | 0.016 | 0.237 | 2.69 | 0.59 |
| Co-$CeO_2$ Ar/$H_2$ | 2.19 | 1.32 | 2.31 | -1.84 | 5.42 | 42.53 |
| Co-$CeO_2$ Air | - | - | 0.049 | 2.175 | 0.72 | 6.81 |
| Y-$CeO_2$ Ar/$H_2$ | 0.2 | 0.12 | 0.169 | 0.741 | 7.18 | 2.35 |
| Y-$CeO_2$ Air | 0.05 | 0.031 | 0.033 | 0.357 | 4.034 | 0.82 |

## Theoretical analysis

The electronic and magnetic properties of $CeO_2$ and TM-doped $CeO_2$ were systematically investigated, focusing on the effect of vacancies, using DFT calculations based on the structure obtained from Rietveld refinement (Fig. 8(a-d)). Some of the results of the Rietveld refinement are presented in the supplementary file, Fig. S8(a, b). All calculations were performed using a $2 \times 2 \times 2$ supercell to mitigate dopant-dopant and vacancy-vacancy interactions. The models were examined: un-doped $CeO_2$, Co-$CeO_2$, $V_O^q$-Co-$CeO_2$, $V_{Ce}$-Co-$CeO_2$, Y-$CeO_2$, $V_O^q$-Y-$CeO_2$, and $V_{Ce}$-Y-$CeO_2$. Here, '$V_O^q$' and '$V_{Ce}$' denote the introduction of an oxygen vacancy with charge state q (e. g. 0, +1, and +2) and cation vacancy from the neighboring site of the TM impurity The DFT calculations were performed using the linear combination of atomic orbitals (LCAO) method, as implemented in QuantumATK software.[79] Local density approximation (LDA) was employed to represent exchange-correlation interactions, utilizing the Perdew-Zunger (PZ) functional within the ATK software.[80–82] HGH pseudopotential was chosen as the basis set for structures. Fermi–Dirac functions were utilized for the occupation method, with a grid mesh cut-off set to 75 Hartree. The k-point mesh was selected in accordance with the LDA exchange-correlation potential. Geometry optimization was carried out using the LBFGS optimizer method, with a self-consistent field iteration of 0.0001 Hartree.[83] During geometry optimization, lattice parameters and atomic positions were optimized until the atomic forces were smaller than 0.05 eV/Å.

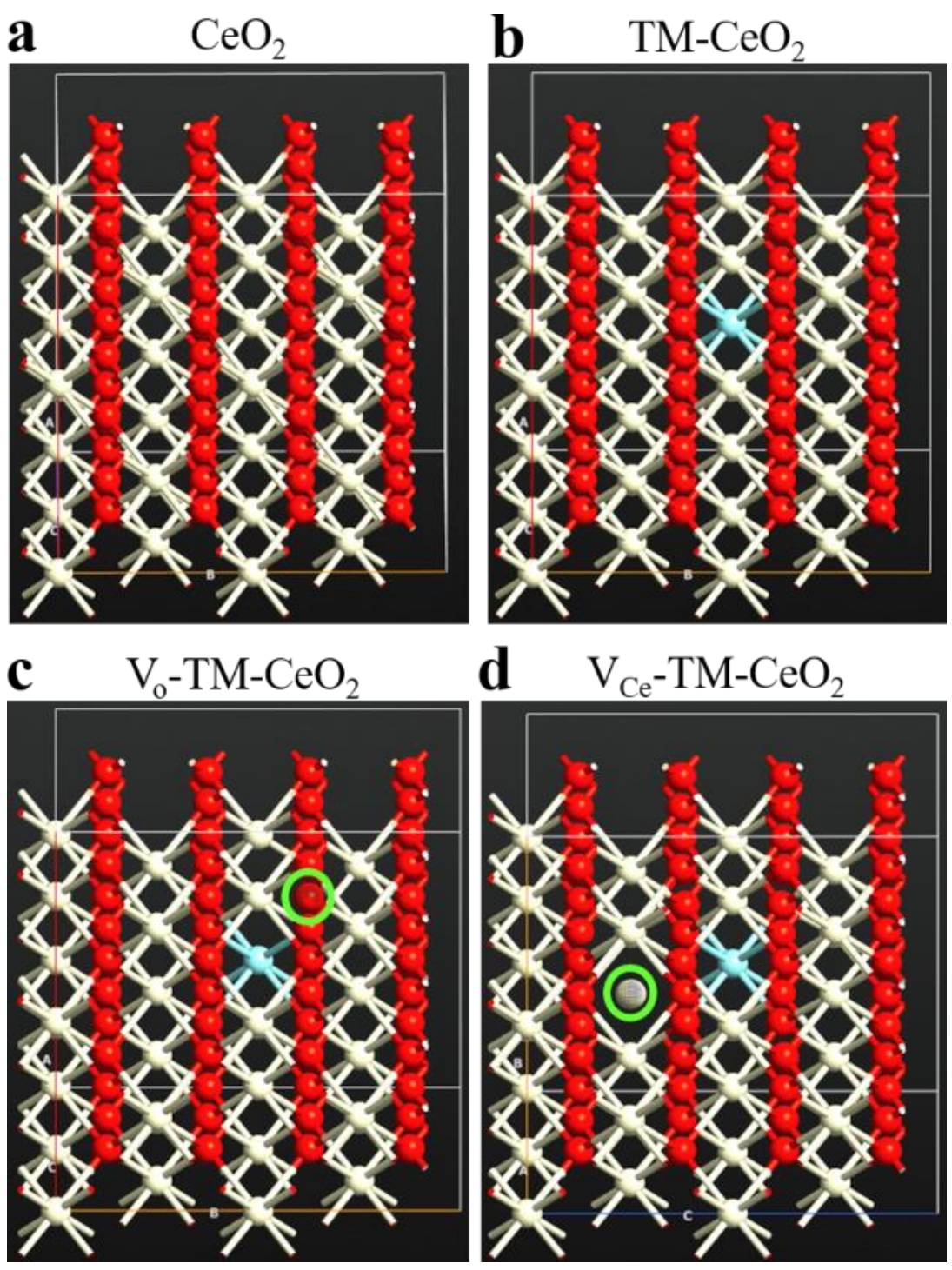

**Figure 8** Crystal structure of (a) $CeO_2$, (b) TM-doped $CeO_2$, (c) Oxygen vacancy in doped $CeO_2$, and (d) Cerium vacancy in doped $CeO_2$. The cream, red, and cyan spheres represent Ce, O, and TM (Co and Y) atoms respectively.

The calculated spin-polarized density of states (DOS) and partial density of states (PDOS) of $CeO_2$ are shown in Fig. 9(a, b). The DOS calculation indicates mirror symmetry in spin-up and spin-down states which confirms the non-magnetic behavior of $CeO_2$, with a band gap of ~2.16 eV existing between O 2p occupied states and Ce 4f unoccupied states.[84] Our DFT calculation for the band gap of un-doped $CeO_2$ gives an underestimated value compared to the experimental value of 3.1 eV. Therefore, the Hubbard U parameter is introduced for bandgap correction.[85] The optimal combination of U = 8 eV for the Ce 4f orbitals and U = 4 eV for the Ce 5d orbitals was found to improve the prediction of the electronic structure. The observed band gap value after the Hubbard correction is approximately 3.1 eV, which is in good agreement with the experimental value, as shown in the DOS and PDOS of $CeO_2$

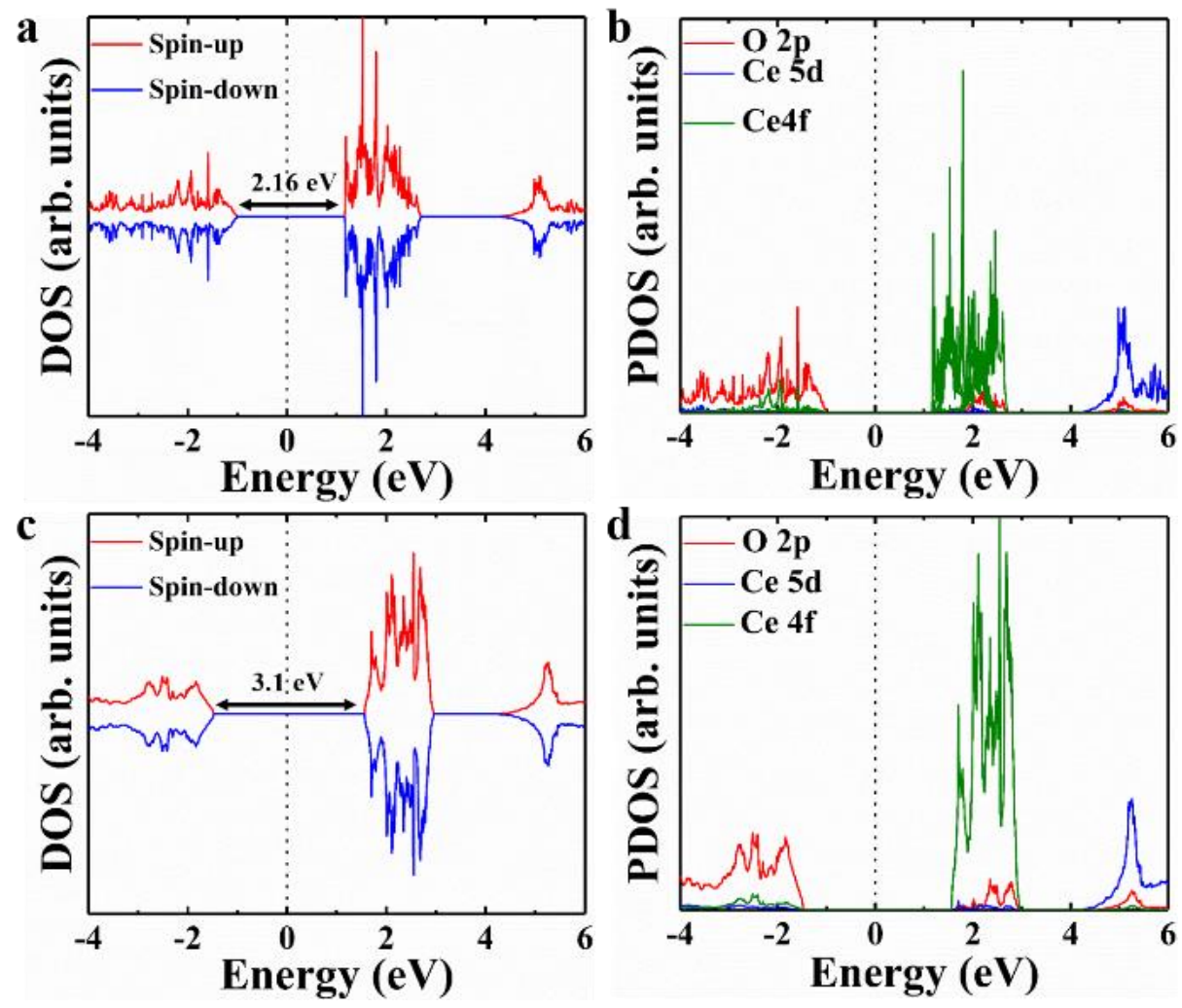

**Figure 9** (a) Spin-polarized density of states and (b) Projected density of states of $CeO_2$ using LDA approach. (c) Spin-polarized density of states and (d) Projected density of states of $CeO_2$ using LDA+U approach.

in Fig. 9(c, d). Further, all calculations are performed using the LDA+U approach.

To evaluate the energy needed to create the neutral vacancies in Co- and Y-$CeO_2$ system, the vacancy formation energy has been calculated by[86]

$$E_{Vac} = E_{defect} - E_{perfect} + E_O/E_{Ce} \quad (12)$$

where, $E_{defect}$ and $E_{perfect}$ are the total energies of the $CeO_2$ and TM-CeO with oxygen or cation vacancy and without vacancy, respectively. The $E_O$ is half of the total energy of oxygen molecules in the gas phase, and $E_{Ce}$ is the total energy per atom of Ce bulk. The calculated results for the oxygen vacancy formation energy are 4.68 eV and 2.76 eV in Co- and Y-doped $CeO_2$, respectively. For the formation of the $V_{Ce}$ vacancy, the energies required are 9.93 eV and 9.41 eV in Co- and Y-doped $CeO_2$, respectively. The positive value of $E_{Vac}$ indicated that energy is needed to form the vacancy.[87] The results indicate that less energy is needed to create $V_O^0$ compared to $V_{Ce}$. In our experimental work, the formation of $V_{Ce}$ by annealing at relatively low temperatures, such as 800°C, is less favorable, while the formation of $V_O^0$ is more favorable. The $V_{Ce}$ can be created through high-energy ion irradiation, so we have not included the effect of $V_{Ce}$ in further calculations.

It is well-known that oxygen vacancies in $CeO_2$ can exist in three charge states: neutral ($V_O^0$), singly charged ($V_O^{+1}$), and doubly charged ($V_O^{+2}$). To investigate the formation energies of different charged states of oxygen vacancies, we used the following formula[88]

$$E_f(V_O^q) = E(V_O^q) - E(V_O^0) + q\,(E_{VBM} + E_F) + E_{corr} \quad (12)$$

Where $E(V_O^q)$ is the total energy of O vacancy with charge q, $E(V_O^0)$ is the total energy of neutral O vacancy, $E_{VBM}$ is the valance band maximum energy, $E_F$ is the Fermi energy, and $E_{corr}$ accounts for the correction due to electrostatic interactions between the background charges and the defects. In this case, the $E_{corr}$ was calculated to be 0.102 eV for the +1 charge state and 0.408 eV for the +2 charge state.[89] The calculated formation energies for oxygen vacancies in different charge states in TM-$CeO_2$ are presented as a function of Fermi energy (shifting from the VBM to the band gap) are shown in Fig. 10 (a, b). The formation energy of the $V_O^0$ is independent of the Fermi level. Additionally, the formation energy of positively charged defects increases as the Fermi energy rises, consistent with expectations for charged defects in semiconductors. In Co-$CeO_2$, the $V_O^{+2}$ is stable at a low Fermi level energy value. After that, a transition occurs from $V_O^{+2}$ to $V_O^{+1}$ (denoted as $\varepsilon^{+2/+1}$) at 0.08 eV. As Fermi energy increases further, the formation energy of $V_O^{+1}$ becomes stable. The $V_O^0$ is not stable state in Co-$CeO_2$ under these conditions. In Y-$CeO_2$, the formation energy of the $V_O^{+2}$ is stable until the Fermi energy reaches 1.4 eV. After this, $\varepsilon^{+2/0}$ transition occurs. This indicates that neutral vacancies become stable as the Fermi level shifts higher.

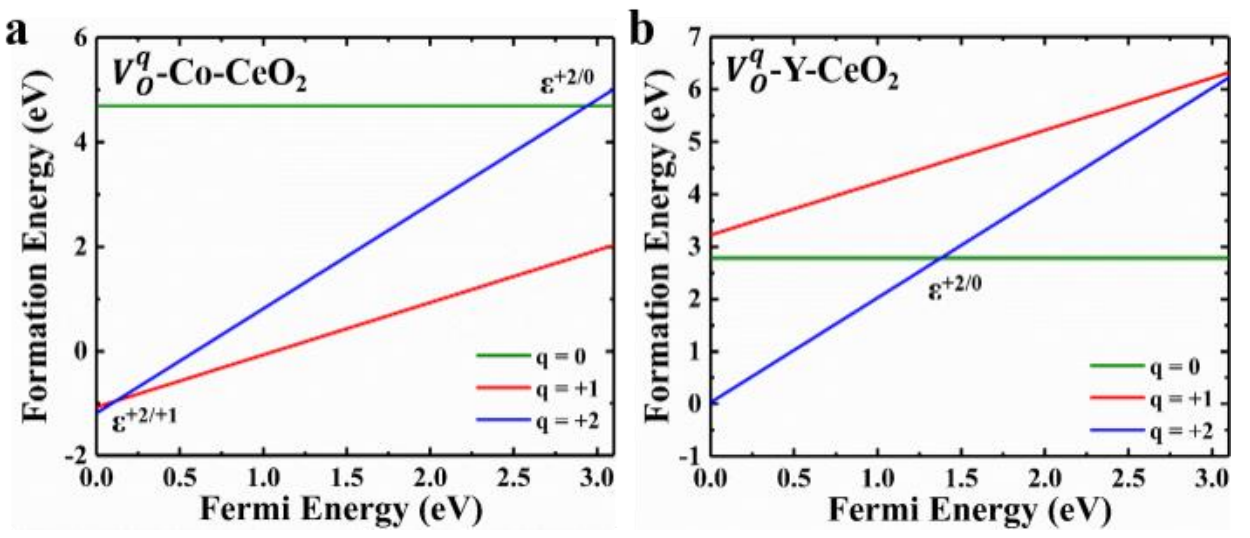

**Figure 10** Formation energy of oxygen vacancy at different charge states in (a) Co-$CeO_2$ and (b) Y-$CeO_2$.

Further, the spin-polarized DOS of TM-$CeO_2$, along with the charged oxygen vacancies ($V_O^q$), are illustrated in Fig. 11 (a-h). All structures, except Y-$CeO_2$, exhibit broken symmetry between the spin-up and spin-down states, confirming the magnetic behavior of these systems.[90] In Co-$CeO_2$, the Co atoms introduce localized gap states, with band gaps of 0.43 eV and 1.36 eV for spin-up and spin-down states, respectively.

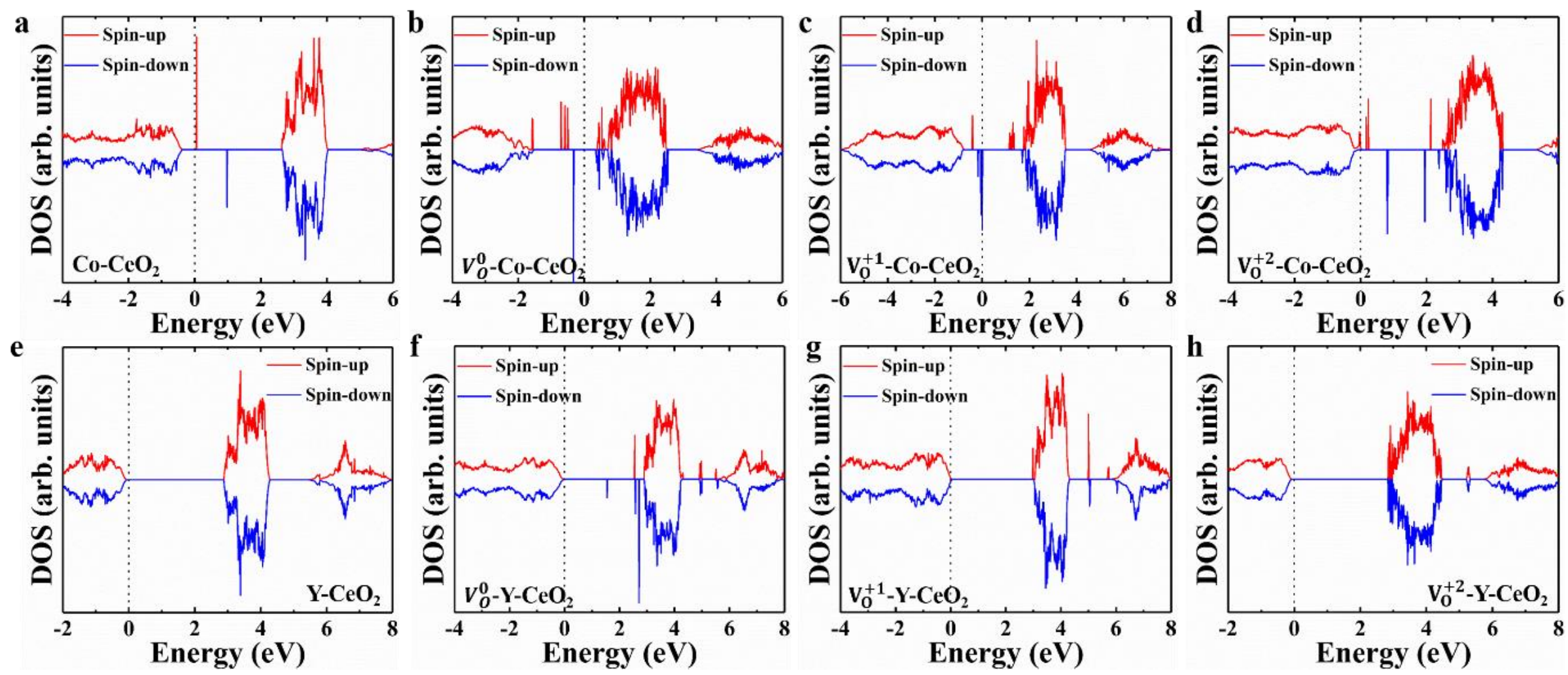

**Figure 11** Spin-polarized density of states of (a) Co-$CeO_2$, (b) $V_O^0$-Co-$CeO_2$, (c) $V_O^{+1}$-Co-$CeO_2$, (d) $V_O^{+2}$-Co-$CeO_2$, (e) Y-$CeO_2$, (f) $V_O^0$-Y-$CeO_2$, (g) $V_O^{+1}$-Y-$CeO_2$, (h) $V_O^{+2}$-Y-$CeO_2$ The Fermi level is shown at the zero of energy.

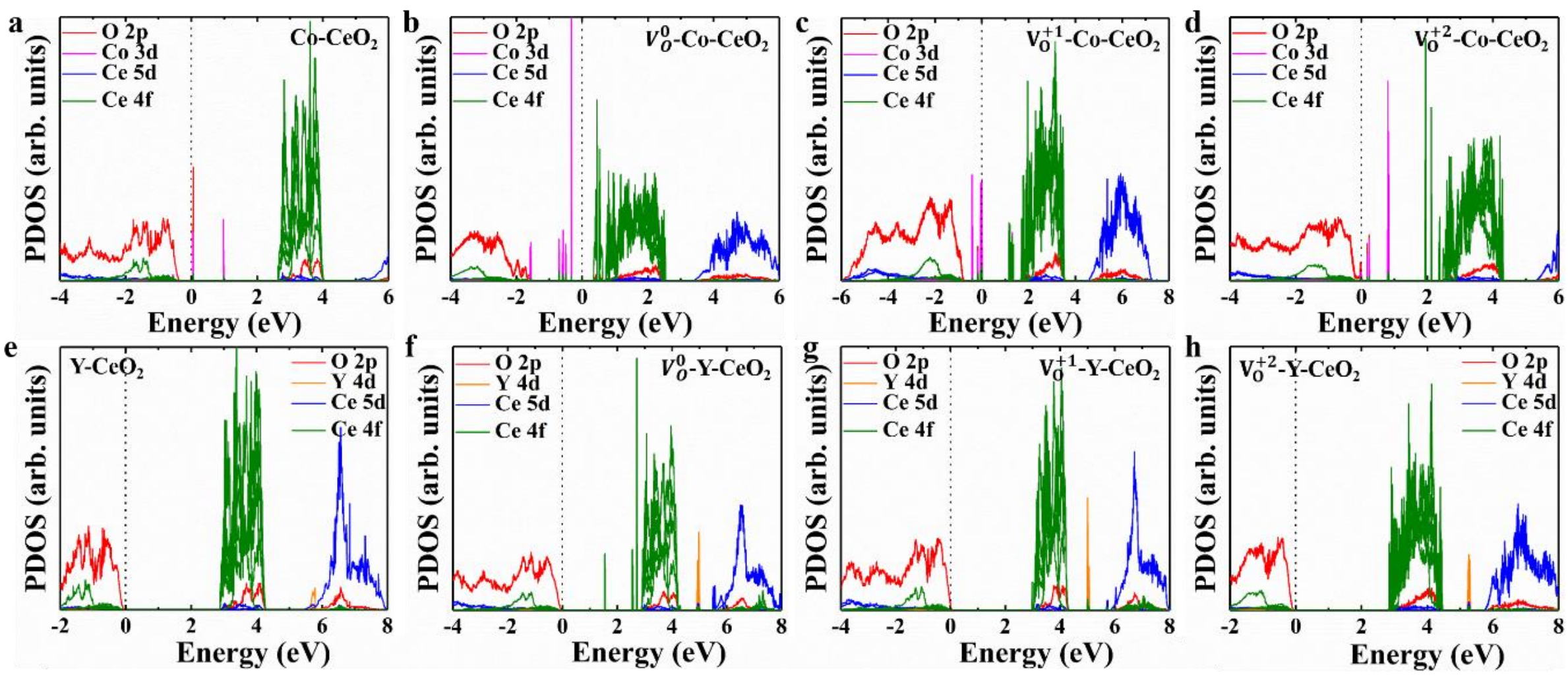

**Figure 12** Spin-polarized partial density of states of (a) Co-CeO$_2$, (b) $V_O^0$-Co-CeO$_2$, (c) $V_O^{+1}$-Co-CeO$_2$, (d) $V_O^{+2}$-Co-CeO$_2$, (e) Y-CeO$_2$, (f) $V_O^0$-Y-CeO$_2$, (g) $V_O^{+1}$-Y-CeO$_2$, (h) $V_O^{+2}$-Y-CeO$_2$. The Fermi level is shown at the zero of energy.

In $V_O$-Co-CeO$_2$, due to the presence of two extra electrons, the dopant states shifted below the Fermi level which indicates the extra electrons are occupied by the dopant states and few states are observed near the conduction band, which can be attributed to the reduction of $Ce^{4+}$ to $Ce^{3+}$. The observed band gaps of spin-up and spin-down states are mentioned in Table 5. In $V_O^{+1}$-Co-CeO$_2$, the Fermi level starts shifting toward the valence band due to a decreased number of electrons, corresponding spin-up and spin-down band gaps listed in Table 5. [91] In $V_O^{+2}$-Co-CeO$_2$, the Fermi level again shifts toward the valence band, and the dopant states lie above the Fermi level. The structure exhibits half-metallic behavior. The PDOS analysis reveals that these impurity states within the band gap mainly arise from the hybridization between Co 3d and O 2p states, with a slight contribution from Ce 4f states as shown in Fig. 12(a). The oxygen vacancy generates $Ce^{3+}$ states near the conduction band, as clearly seen in Fig. 12(b-d). The DFT calculations show multiple donor and $Ce^{3+}$ states originating within the band gap for Co-CeO$_2$ and $V_O^q$-Co-CeO$_2$, which reduces the band gap and is consistent with experimental data of Co-CeO$_2$. Further, the spin-polarized DOS of Y-CeO$_2$ exhibited non-magnetic p-type semiconductor behavior with a band gap of 2.97 eV (Fig. 11(e)). The PDOS revealed that Y impurities introduce states between the Ce 4f and Ce 5d states, mainly arising from the hybridization of Y 4d and Ce 4f states, as shown in Fig. 12(e). In contrast, the spin-polarized DOS of $V_O^0$-Y-CeO$_2$ indicates magnetic behavior, with some spin states near the conduction band corresponding to $Ce^{3+}$ sites, as shown in Fig. 11(f). This is further confirmed by PDOS calculations, which indicate that these states correspond to Ce 4f states (Fig. 12(f)). Moreover, when oxygen vacancies with charge states $V_O^{+1}$ and $V_O^{+2}$ are created in Y-CeO$_2$, the deep-lying Ce states are removed, with a few states remaining near the conduction band as shown in Figs. 11 (g, h) and 12 (g, h). This behavior is caused by the removal of extra electrons. In $V_O^q$-Y-CeO$_2$, a slight change in the band gap is observed, which is in alignment with experimental data for Y-$CeO_2$. However, no major shift in the Fermi level is observed in all $V_O^q$-Y-$CeO_2$ structures, indicating p-type semiconductor nature.

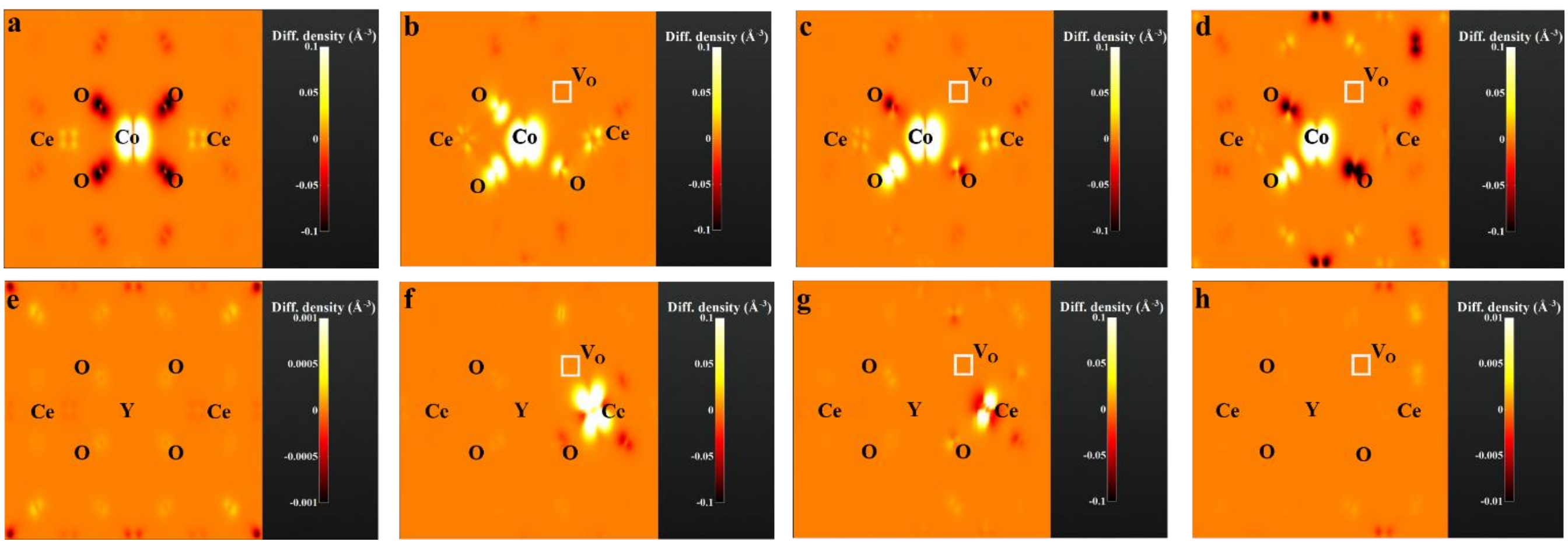

Figure 13 Spin-Charge difference density along (110) plane (a) Co-CeO$_2$, (b) $V_O^0$-Co-CeO$_2$, (c) $V_O^{+1}$-Co-CeO$_2$, (d) $V_O^{+2}$-Co-CeO$_2$, (e) Y-CeO$_2$, (f) $V_O^0$-Y-CeO$_2$, (g) $V_O^{+1}$-Y-CeO$_2$, (h) $V_O^{+2}$-Y-CeO$_2$, the oxygen vacancy are denoted by square box. This is obtained by taking the difference between the charge densities of spin-up and spin-down ($\rho = \rho\uparrow - \rho\downarrow$).

**Table 5.** Spin-polarized band gaps and Magnetic moment of un-doped, Co- and Y-doped $CeO_2$ with and without oxygen vacancy.

| | Band gap (eV) | | Magnetic Moment/supercell ($\mu_B$) |
|---|---|---|---|
| | Spin-up | Spin-down | |
| $CeO_2$ (LDA) | 2.16 | 2.16 | 0 |
| $CeO_2$ | 3.1 | 3.1 | 0 |
| Co-$CeO_2$ | 0.43 | 1.37 | 1 |
| $V_O^0$-Co-$CeO_2$ | 0.89 | 0.68 | 3 |
| $V_O^{+1}$-Co-$CeO_2$ | 1.56 | 0.06 | 2 |
| $V_O^{+2}$-Co-$CeO_2$ | 0 | 0.89 | 1.4 |
| Y-$CeO_2$ | 2.97 | 2.97 | 0 |
| $V_O^0$-Y-$CeO_2$ | 2.6 | 1.57 | 2 |
| $V_O^{+1}$-Y-$CeO_2$ | 2.92 | 2.98 | 1 |
| $V_O^{+2}$-Y-$CeO_2$ | 2.91 | 2.91 | 0 |

To further analyze the effect of charged O vacancies and doping on magnetic properties, the calculated magnetic moment listed in Table 5 and spin-charge difference density along the [110] plane of all structures are shown in Fig. 13(a-h).[92] In Co-$CeO_2$, the resulting net magnetic moment is 1 $\mu_B$ per supercell. Notably, the induced magnetic moment is contributed by the Co atom (1.53 $\mu_B$), O atoms (-0.1 $\mu_B$), and Ce atoms (0.26 $\mu_B$). The spin-charge difference density calculation, shown in Fig. 13(a), reveals that Co 3d electrons cause polarization of the surrounding O and Ce atoms. This suggests an AFM coupling between the doped Co atom and neighboring O atoms, whereas FM coupling occurs between the doped Co atom and neighboring Ce atoms. The magnetic moment of the Co (1.53 $\mu_B$) suggests that the Co dopant does not fully oxidize to the $Co^{2+}$ state. Further, in $V_O^0$-Co-$CeO_2$, the net moment increases to 3 $\mu_B$ per supercell. The induced magnetic moment is contributed by the Co atom (2.49 $\mu_B$), O atoms (0.17 $\mu_B$), and Ce atoms (0.37 $\mu_B$). This enhancement in the moment is due to the presence of two unbound electrons created by the neutral $V_O$, and this is majorly localized on Co ions and slightly distributed on neighboring Ce ion which reduced $Ce^{4+}$ to $Ce^{3+}$.[93] Here, the magnetic moment on the Co (2.49 $\mu_B$) indicates that the Co dopant is fully oxidized to the $Co^{2+}$ state.[94] The spin alignment is showing FM behavior as shown in Fig. 13(b). On creating charged vacancies ($V_O^{+1}$ and $V_O^{+2}$), the total moment decreases to 2 $\mu_B$ and 1.4 $\mu_B$, respectively, due to the reduction of residual electrons by the charged O vacancies and the major charge is localized on Co ion can be seen from Fig. 13 (c, d). The Y-$CeO_2$ exhibits non-magnetic behavior, resulting in zero magnetic moments, due to the neutral ion behavior of Y in $CeO_2$, with negligible moments from both Ce and O atoms (see Fig. 13(e)). This 0 $\mu_B$ moment of Y suggesting that the Y dopant is present in $Y^{3+}$ state. The formation of a neutral oxygen vacancy in Y-$CeO_2$ ($V_O^0$-Y-$CeO_2$) results in a net magnetic moment of 2 $\mu_B$ per supercell.[92] The spin-charge difference density and magnetic moment analysis show that the electrons localizing on the neighboring Ce atoms contribute a magnetization of 2 $\mu_B$. The contribution of O atoms negligible, as shown in Fig. 13(f). Further, the spin-polarized difference density of $V_O^{+1}$-Y-$CeO_2$ indicates a net moment of 1 $\mu_B$ and the creation of $V_O^{+2}$-Y-$CeO_2$ leads to a decrease in the net magnetic moment of the structure to 0 $\mu_B$, showing that the FM coupling is weakened as the charge state of the vacancy increases, shown in Fig. 13(g, h).

Furthermore, when two TM ions are introduced into $CeO_2$ to study the effect of magnetic coupling between the two TM dopants via $O^{2-}$ and $V_O^q$, the results indicate that the interaction between two Co atoms through $O^{2-}$ and $V_O^q$ shows stable FM coupling. The magnetic moment and energy difference between AFM and FM states are presented in Table 6. In the Co-$V_O^0$-Co system, the FM state is found to be more stable compared to others, indicating that the superexchange interaction in Co-$CeO_2$ is primarily responsible for the FM coupling. In the case of two Y atoms doped in $CeO_2$, the spin un-polarized state is stable in all cases because of the non-magnetic behavior of the Y ion.

**Table 6.** Vacancy formation energy, Band gaps, Magnetic moment, and nature of un-doped, Co- and Y-doped $CeO_2$ with and without oxygen vacancy.

| | $E_{AFM}$-$E_{FM}$ (eV) | Magnetic Moment/supercell ($\mu_B$) |
|---|---|---|
| Co-O-Co | 0.07 | 2.0 |
| Co-$V_O^0$-Co | 4.35 | 7.9 |
| Co-$V_O^{+1}$-Co | 0.03 | 3.2 |
| Co-$V_O^{+2}$-Co | 0.02 | 2.6 |

The above calculations show that in Co-$CeO_2$, the formation of $V_O^{+1}$ is more stable, and thus, magnetic coupling favors the F-center mechanism. The interaction between two Co ions through oxygen ions enhances the FM interaction. Additionally, the presence of $Ce^{3+}$ ions contributes significantly to the magnetic properties, which results in a high level of magnetism in Co-$CeO_2$. In contrast, in Y-doped $CeO_2$, the $V_O^{2+}$ state is more stable because, in our calculations, the Fermi level lies near the valence band edge, indicating non-magnetic behavior. The interaction between two Y ions through oxygen ions, along with the $V_O^q$ states also show non-magnetic behavior. Therefore, in Y-$CeO_2$, the possible mechanism for FM interaction is primarily attributed to the formation of $Ce^{3+}$ ions.

## Conclusions

Dopants and annealing environments are the critical parameters in tuning the defect chemistry and consequently the magnetic properties of DMOs. The present work concludes that annealing in an Ar/$H_2$ atmosphere induces more vacancies compared to air annealing. The vacancies play a crucial role in modulating the magnetic properties of both un-doped and TM-doped $CeO_2$. Pure $CeO_2$ exhibited weak RTFM, whereas doped samples showed enhanced magnetization, with Co-$CeO_2$ displaying higher magnetization than Y-$CeO_2$. Ar/$H_2$ annealed samples exhibited higher saturation and residual magnetization compared to air-annealed samples. The DFT calculations supported these experimental findings, showing a reduction in the band gap of Co-$CeO_2$ due to hybridized states of the dopant and vacancies, while Y-$CeO_2$ showed minimal change in the band gap. Additionally, charged oxygen vacancies in doped $CeO_2$ provide a deeper understanding of the origin of magnetism. In Co-$CeO_2$, the localized magnetic moment on the Co ion, coupled with stable $V_O^{+1}$ states and superexchange interactions play a crucial role in the strong FM coupling. In contrast, in Y-$CeO_2$, only the formation of $Ce^{3+}$ ions are responsible for the observed magnetism. Present findings have significant implications for the application of DMOs in spintronics, photonics, and catalysis, suggesting that controlled annealing processes can optimize their magnetic performance.

## Author contributions

Hemant Arora: Conceptualization; Data curation; Formal analysis; Writing original draft, review & editing. Atul Bandyopadhyay: Conceptualization; Data curation; Formal analysis; review & editing. Arup Samanta: Conceptualization; Formal analysis; Funding acquisition; Resources; Supervision; Review & editing.

## Conflicts of interest

The authors declared that they have no known competing financial interests or personal relationships that could have appeared to influence the work reported in this paper.

## Acknowledgements

The authors thank Prof. Sparsh Mittal and Prof. Sanjeev Manhas for providing computational resources under Project No. ECR/2017/000622 and MIT-896-ECD EICT, respectively. Partial funding for this work came from IISc-MHRD (Project No. STARS-2/2023-0715) and DST (Project No. SR/FST/PS-II/2019/84) in India. H.A. also appreciates the research scholarship provided by the UGC, India.

# Supplementary Information

## Annealing-Induced Magnetic Modulation in Co- and Y-doped $CeO_2$: Insights from Experiments and DFT

Hemant Arora[1], Atul Bandyopadhyay[2,*], and Arup Samanta[1,3,*]

[1]Quantum/Nano Science and Technology Laboratory, Department of Physics, Indian Institute of Technology Roorkee, Roorkee-247667, Uttarakhand, India

[2]Department of Physics, University of Gour Banga, Mokdumpur, Malda, West Bengal 732103, India

[3]Centre for Nanotechnology, Indian Institute of Technology Roorkee, Roorkee-247667, Uttarakhand, India

* Corresponding author. Email address: atulbandyopadhyay@yahoo.com; arup.samanta@ph.iitr.ac.in

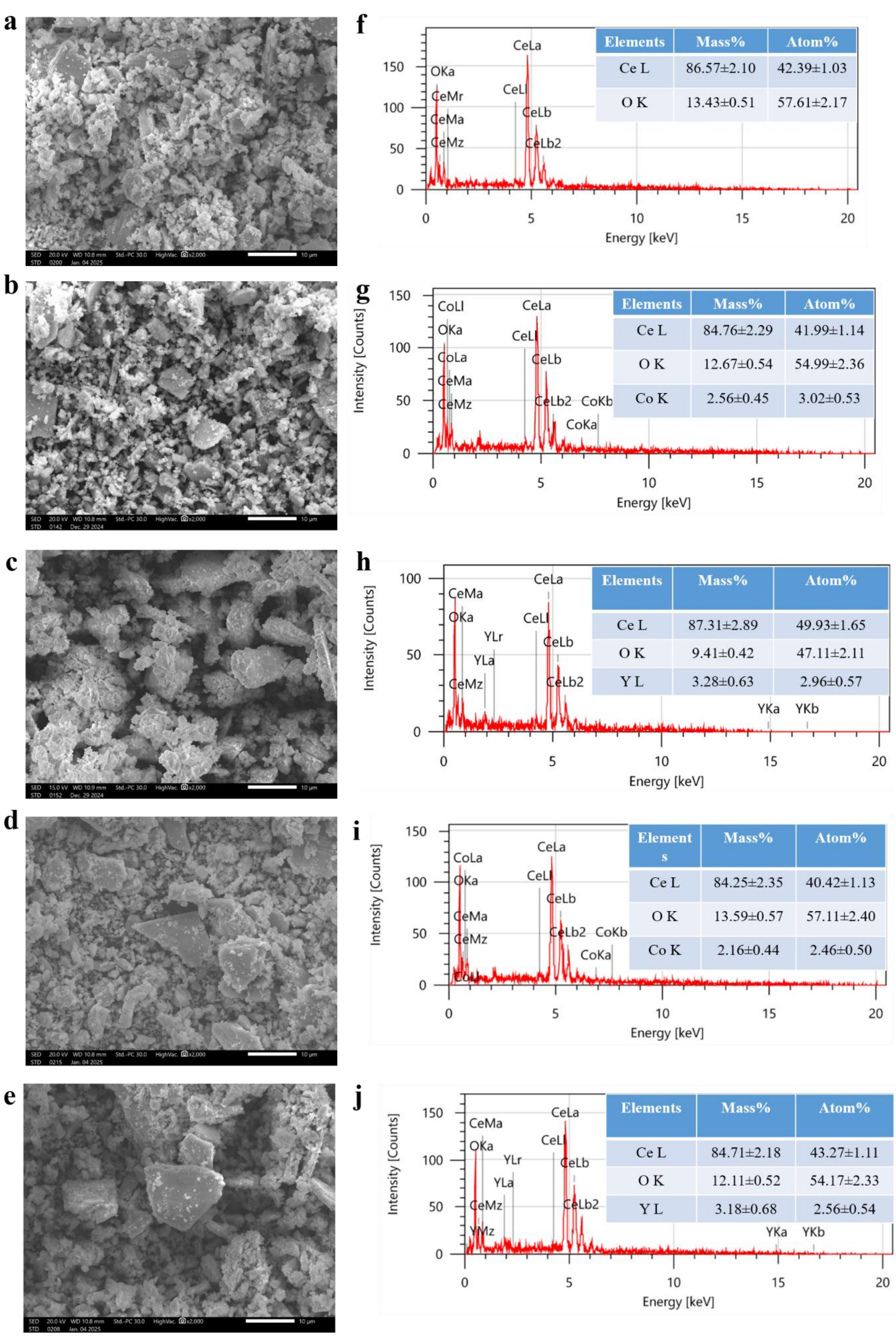

(f)

| Elements | Mass% | Atom% |
|---|---|---|
| Ce L | 86.57±2.10 | 42.39±1.03 |
| O K | 13.43±0.51 | 57.61±2.17 |

(g)

| Elements | Mass% | Atom% |
|---|---|---|
| Ce L | 84.76±2.29 | 41.99±1.14 |
| O K | 12.67±0.54 | 54.99±2.36 |
| Co K | 2.56±0.45 | 3.02±0.53 |

(h)

| Elements | Mass% | Atom% |
|---|---|---|
| Ce L | 87.31±2.89 | 49.93±1.65 |
| O K | 9.41±0.42 | 47.11±2.11 |
| Y L | 3.28±0.63 | 2.96±0.57 |

(i)

| Elements | Mass% | Atom% |
|---|---|---|
| Ce L | 84.25±2.35 | 40.42±1.13 |
| O K | 13.59±0.57 | 57.11±2.40 |
| Co K | 2.16±0.44 | 2.46±0.50 |

(j)

| Elements | Mass% | Atom% |
|---|---|---|
| Ce L | 84.71±2.18 | 43.27±1.11 |
| O K | 12.11±0.52 | 54.17±2.33 |
| Y L | 3.18±0.68 | 2.56±0.54 |

**Figure S1** SEM images of (a) $CeO_2$, (b) Co-$CeO_2$ Ar/$H_2$, (c) Y-$CeO_2$ Ar/$H_2$, (d) Co-$CeO_2$ Air, and (e) Y-$CeO_2$ Air. EDS spectrum of (f) $CeO_2$, (g) Co-$CeO_2$ Ar/$H_2$, (h) Y-$CeO_2$ Ar/$H_2$, (i) Co-$CeO_2$ Air, and (j) Y-$CeO_2$ Air.

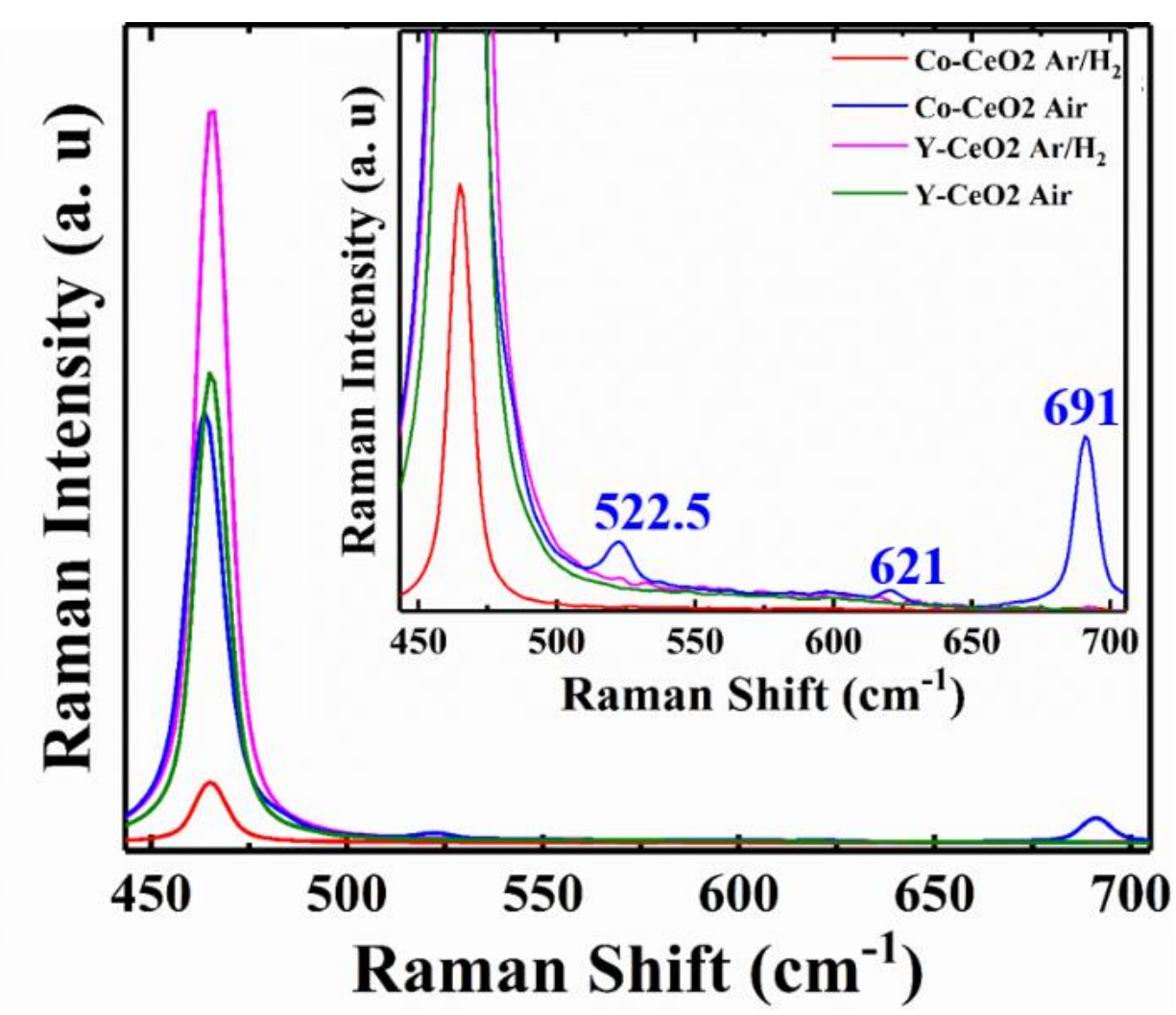

**Figure S2** Room-temperature Raman spectra of Co and Y-$CeO_2$ annealed in Ar/$H_2$ and Air environment, respectively.

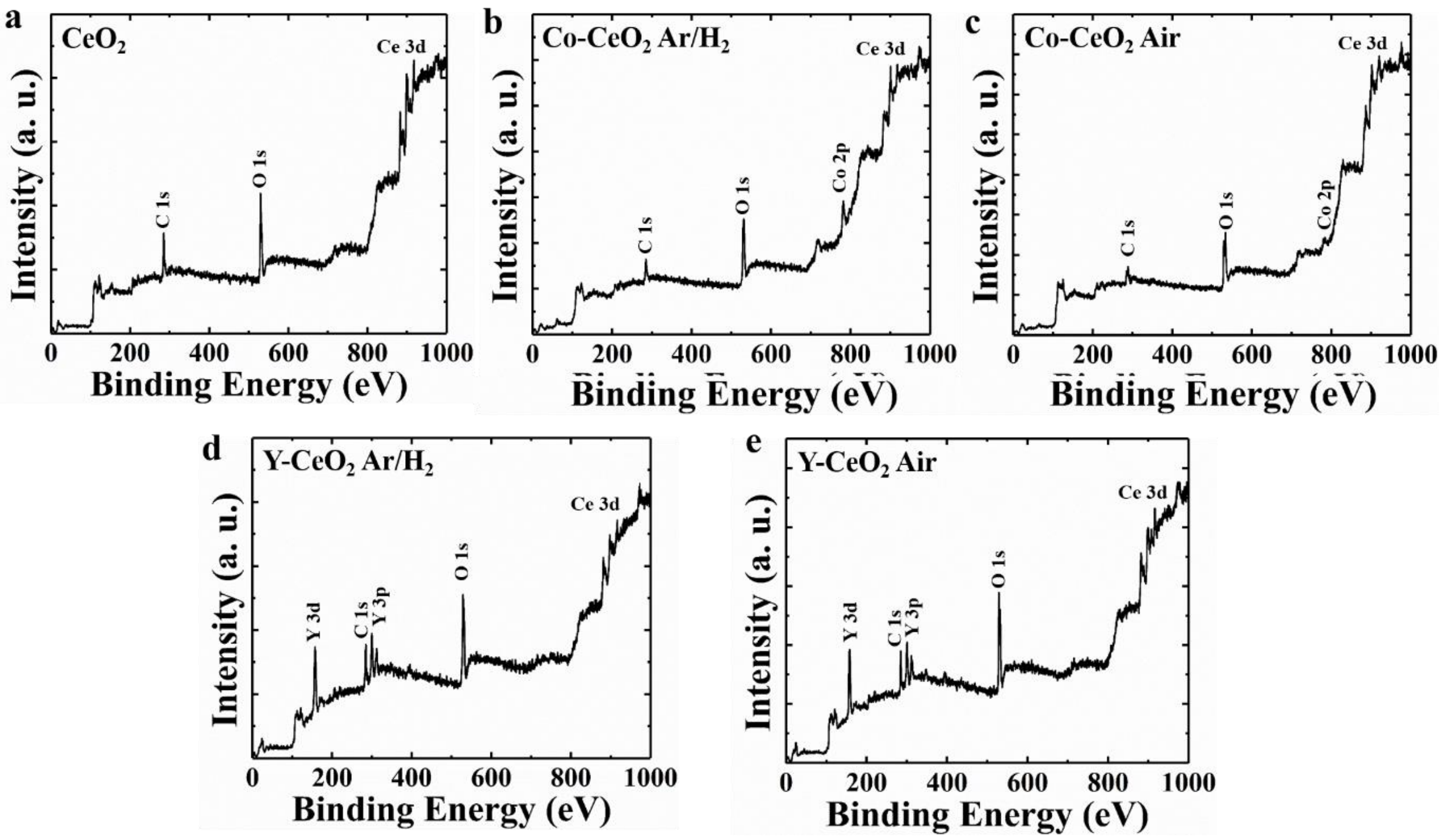

**Figure S3** The XPS survey spectrum of (a) $CeO_2$, (b) Co-$CeO_2$ Ar/$H_2$, (c) Co-$CeO_2$ Air (d) Y-$CeO_2$ Ar/$H_2$, and (e) Y-$CeO_2$ Air.

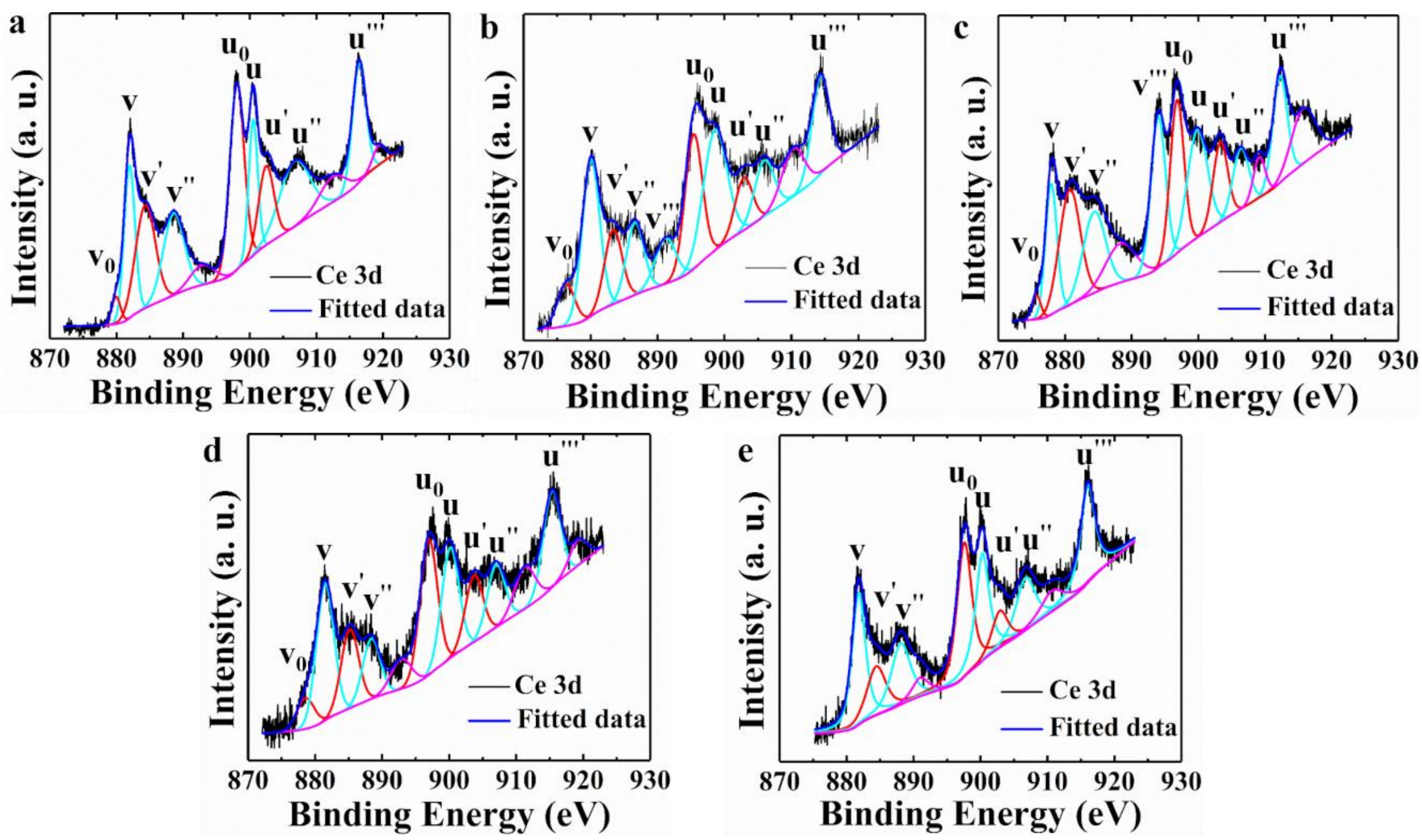

**Figure S4** Ce 3d spectra of (a) $CeO_2$, (b) Co-$CeO_2$ Ar/$H_2$, (c) Co-$CeO_2$ Air (d) Y-$CeO_2$ Ar/$H_2$, and (e) Y-$CeO_2$ Air.

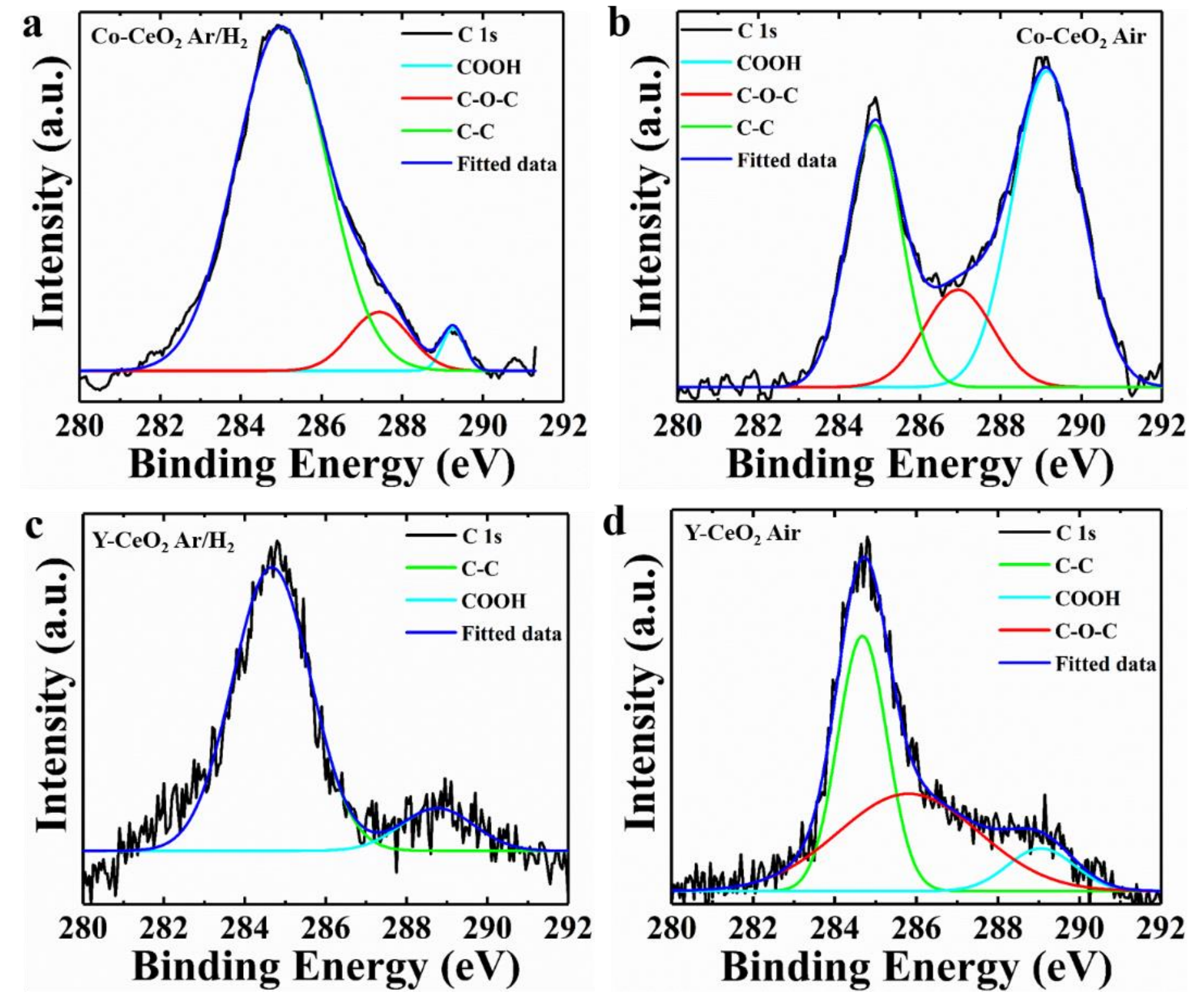

**Figure S5** C 1s spectra of (a) Co-$CeO_2$ Ar/$H_2$, and (b) Co-$CeO_2$ Air, (c) Y-$CeO_2$ Ar/$H_2$, and (d) Y-$CeO_2$ Air.

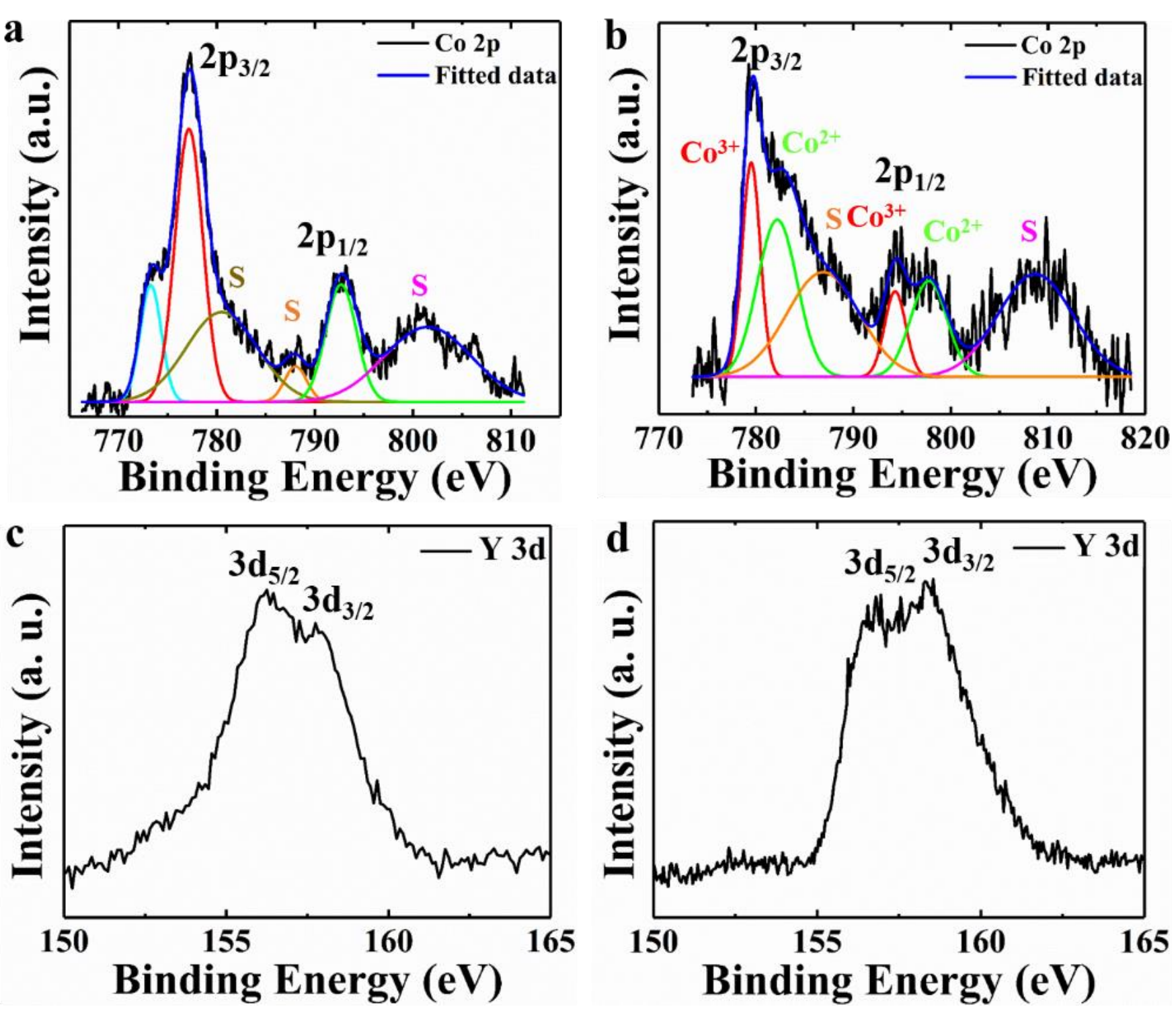

**Figure S6** Co 2p spectra of (a) Co-$CeO_2$ Ar/$H_2$, and (b) Co-$CeO_2$ Air. Y 3d spectra of (c) Y-$CeO_2$ Ar/$H_2$, and (d) Y-$CeO_2$ Air. The S denotes the satellite peak.

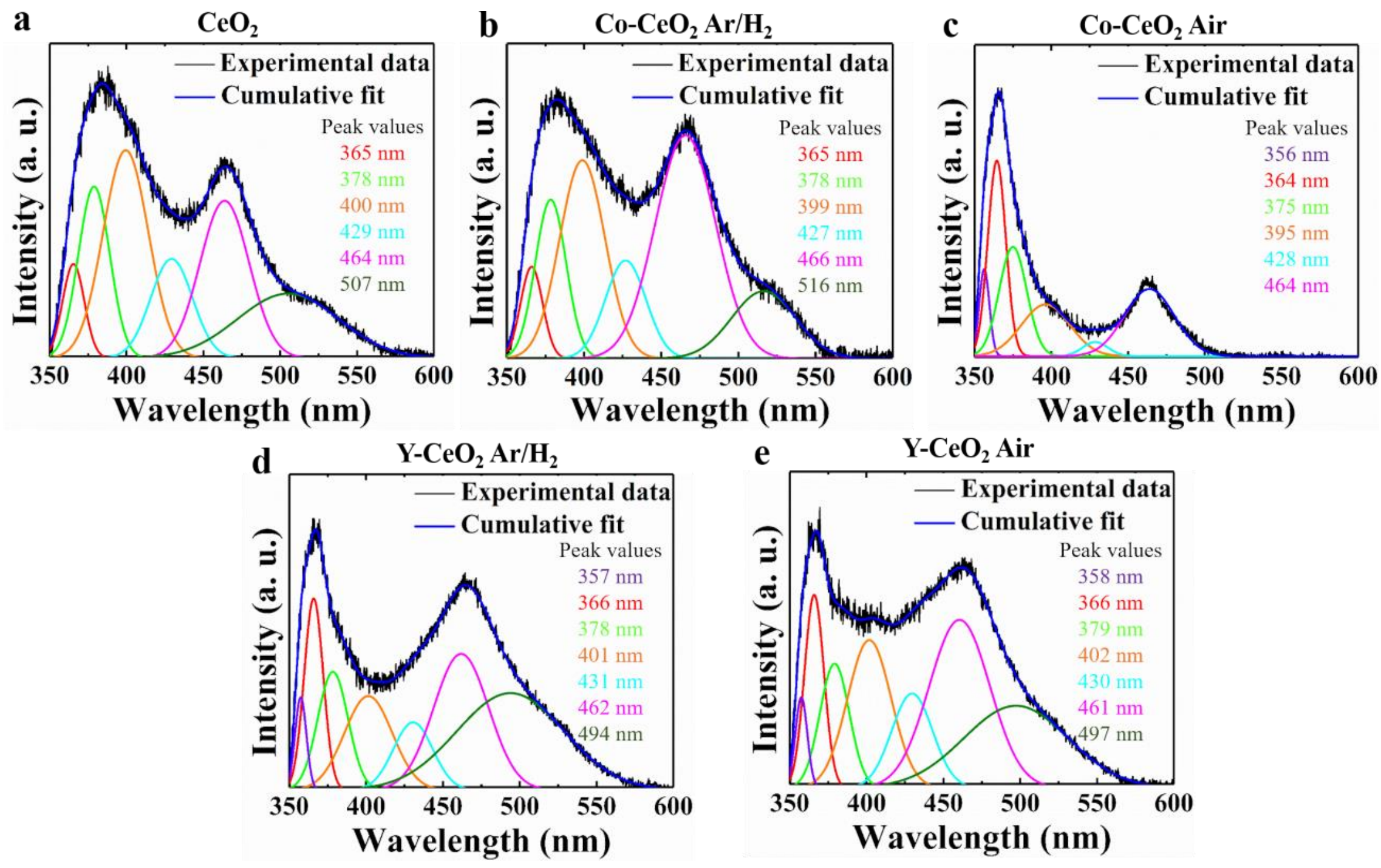

**Figure S7** Photoluminescence spectra of (a) $CeO_2$ (b) Co-$CeO_2$ Ar/$H_2$, (c) Co-$CeO_2$ Air, (d) Y-$CeO_2$ Ar/$H_2$, (e) Y-$CeO_2$ Air.

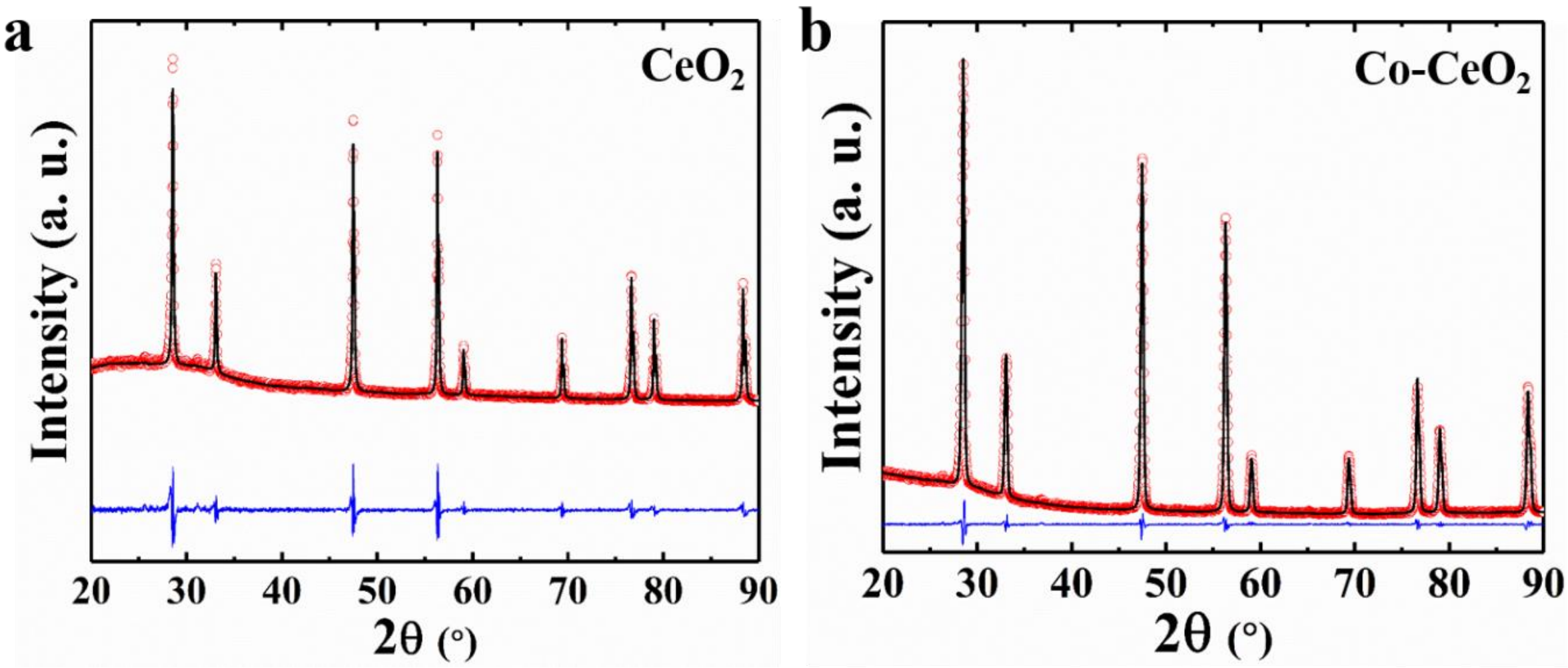

**Figure S8** Rietveld refinement profile of XRD data of (a) $CeO_2$ and (b) Co-$CeO_2$.